\documentclass[twocolumn]{emulateapj}
\shorttitle{Line Profiles from Filaments}
\shortauthors{Smith et al.}

\newcommand{\fig}{Figure }
\newcommand{\sect}{Section }
\newcommand{\tab}{Table }

\newcommand{\msun}{\,M$_{\odot}$ }

\newcommand{\cmc}{\,cm$^{-3}$}
\newcommand{\cmsq}{\,cm$^{-2}$}
\newcommand{\kms}{\,kms$^{-1}$}

\newcommand{\E}{\times 10}
\newcommand{\degree}{$^{\circ}$}
\newcommand{\n}{N$_2$H$^+$}

\begin{document}

\title{Line Profiles of Cores within Clusters: I. The Anatomy of a Filament}

\author{Rowan J. Smith and Rahul Shetty}
\affil{Zentrum f\"ur Astronomie der Universit\"at Heidelberg, Institut f\"ur Theoretische Astrophysik, Albert-Ueberle-Str. 2, 69120 Heidelberg, Germany}
\email{rowanjsmith.astro@googlemail.com}
\author{Amelia M. Stutz}
\affil{Max-Planck-Institut f\"ur Astronomie, K\"onigstuhl 17, D-69117 Heidelberg, Germany}
\and
\author{Ralf S. Klessen}
\affil{Zentrum f\"ur Astronomie der Universit\"at Heidelberg, Institut f\"ur Theoretische Astrophysik, Albert-Ueberle-Str. 2, 69120 Heidelberg, Germany}

\begin{abstract}
Observations are revealing the ubiquity of filamentary structures in molecular clouds. As cores are often embedded in filaments, it is important to understand how line profiles from such systems differ from those of isolated cores. We perform radiative transfer calculations on a hydrodynamic simulation of a molecular cloud in order to model  line emission from collapsing cores embedded in filaments. We model two optically thick lines, CS(2-1) and HCN(1-0), and one optically thin line, \n(1-0), from three embedded cores. In the hydrodynamic simulation, gas self-gravity, turbulence, and bulk flows create filamentary regions within which cores form. Though the filaments have large dispersions, the \n(1-0) lines indicate subsonic velocities within the cores. We find that the observed optically thick line profiles of CS(2-1) and HCN(1-0) vary drastically with viewing angle. In over 50\% of viewing angles, there is no sign of a blue asymmetry, an idealised signature of infall motions in an isolated spherical collapsing core. Profiles which primarily trace the cores, with little contribution from the surrounding filament, are characterised by a systematically higher HCN(1-0) peak intensity. The \n(1-0) lines do not follow this trend. We demonstrate that red asymmetric profiles are also feasible in the optically thick lines, due to emission from the filament or one-sided accretion flows onto the core. We conclude that embedded cores may frequently undergo collapse without showing a blue asymmetric profile,
and that observational surveys including filamentary regions may underestimate the number of collapsing cores if based solely on profile shapes of optically thick lines. 
\end{abstract}



\section{Introduction}\label{intro}
Star-forming regions are complex systems observed projected onto the plane of the sky. Therefore it has always been a difficult task to determine the true geometries and dynamics within them. One key tool for disentangling the true dynamical state of star-forming cores is the blue infall asymmetry. This effect \citep{Zhou91,Zhou92,Walker94,Myers96} relies on the fact that within a collapsing core there is both a centrally increasing density profile, and two points along a line of sight with the same velocity. In optically thick species ($\tau>1$) only the foreground gas at a given velocity is visible. This leads to an asymmetry in the double-peaked emission line profile as the visible blue emission from the far side of the core originates from higher density gas. The blue infall asymmetry was derived under the assumption of spherical symmetry. 

However, observations increasingly show that this is an ideal rarely achieved in molecular clouds. For instance, \citet{Andre10} show that star-forming cores are frequently part of filaments, \citet{Bontemps10} observed that massive star-forming regions contain multiple substructures, and \citet{Tobin10} and \citet{Stutz09} found that asymmetries can be found on scales as small as 1000 AU. In this work we will revisit the asymmetric collapse profile using simulations of star cluster formation within a large molecular cloud. This paper examines the case of irregular cores embedded within filaments, and future work will consider the case of massive star-forming regions.

\citet{Zhou92} calculated the line profiles expected to arise from an inside-out \citet{Shu77} collapse and from Larson-Penston collapse \citep{Larson69,Penston69}. Both models produced an asymmetric collapse profile, and the resulting line-widths could be used to distinguish between the two collapse profiles. A follow up study \citep{Zhou93} showed that the line profile of the near-spherical Bok globule B335 matched their expected Shu collapse profile. This result was initally confirmed by \citet{Choi95}, but later studies have shown that more complex models may actually be needed \citep[e.g.][]{Wilner00,Stutz08}. Other variants of collapse models have been studied. \citet{Walker94} carried out line modelling of prolate collapsing cores, not just spheres, and once again found blue asymmetric line profiles. \citet{Myers96} used a simple analytic model of two converging layers of gas and produced the same result. The study of spherical core line profiles has continued in more recent years. In particular \citet{Rawlings01}, \citet{Tsamis08} and \citet{Stahler10} have shown the importance of accurately modelling the chemical abundances within cores, particularly since optically thick carbon bearing species, such as CO and CS, freeze out within the dense core centres \citep{Tafalla02}.

Are these models sufficiently accurate to capture the behaviour of realistic star-forming cores? One potential difficulty is that for cores embedded at the end of a collapsing filament or other complex structure, accretion onto the core may predominantly occur from just one direction, as seen in \citet{Smith11a}. This may lead to either a blue or red asymmetry depending on which side of the core the accretion flow is located. Similarly, more complex morphologies and internal velocities within the filament and surrounding envelope could lead to self-absorption in the envelope obscuring the velocities from the core. Investigations into numerical simulations have demonstrated that the structures identified in synthetic maps or position-position-velocity cubes may not necessarily correspond to contiguous structures in three dimensional space, due to the geometric projection along the line of sight (e.g. \citealt{Pichardo00, Ostriker01, Shetty10}). 

Cores embedded within filaments are a common feature of molecular clouds, as shown by recent \textit{Herschel} observations \citep{Arzoumanian11,Menshchikov10,Andre10}. Moreover, in \citet{Smith11a} we specifically studied the geometry of the smallest scales of collapsing `Class 0' cores from a molecular cloud simulation, and found that even on scales of a $\sim 1000$ AU, the cores were irregular, extended structures. The only way to understand the effects of a filamentary geometry on resulting emission line profiles is to directly test these scenarios with radiative transfer line modelling.

Despite the many complicating factors, blue asymmetric collapse profiles are often observed. \citet{Gregersen97} observed a sample of 23 Class 0 sources and found blue asymmetric line profiles in 9 cases, compared to 3 sources with a red asymmetry. Similarly, `blue' cores have been found in surveys by \citet{Lee99}, \citet{Gregersen00}, \citet{Wu03} and \citet{Fuller05}. On the other hand, while these surveys find potential infall candidates, the majority of the cores do not have a blue asymmetric profile. The question is, could a greater fraction of cores be collapsing than what is inferred by blue asymmetric line profiles?

We focus on the line profiles arising from collapsing cores embedded within the irregular filaments from which they are formed. The layout of this paper is as follows. In \sect \ref{theory} we describe in more detail the physical effects which lead to a blue-peaked asymmetric line profile. In \sect \ref{model} we describe our methods and models, and  in \sect \ref{results} we outline our results, paying particular attention to the viewing angle of the model. In \sect \ref{discussion} we discuss how our results may aid interpretation of observations. Finally in \sect \ref{conclusions} we summarise our conclusions.

\section{An asymmetric line profile}\label{theory}

\begin{figure}
\begin{center}
\includegraphics[width=2.7in]{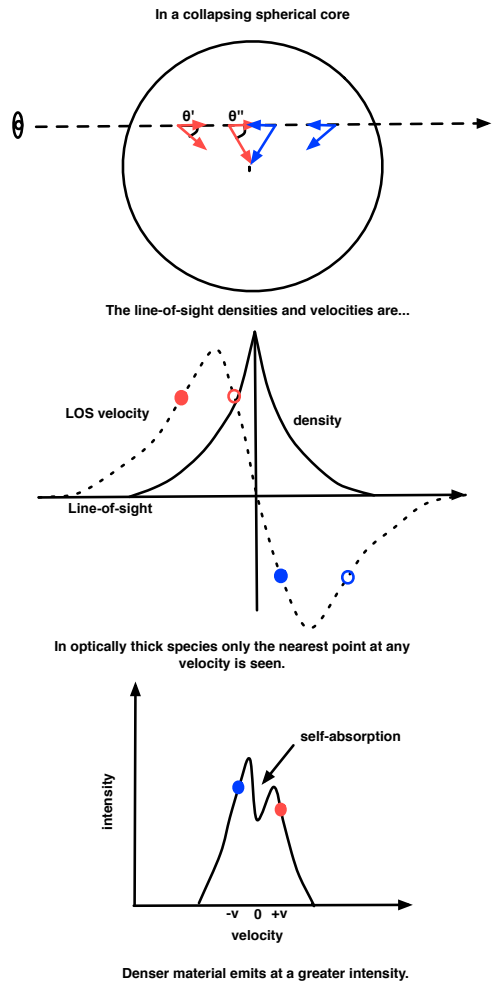}
\caption{A schematic showing the origin of a blue asymmetric line profile from a spherically symmetric collapsing core.}
\label{schematic}
\end{center}
\end{figure}

\fig \ref{schematic} shows a schematic describing the origin of the blue line asymmetry in the simple case of spherical collapsing core (see also \citealt{Evans99}). The gas falls radially towards the centre, however, we only observe the velocity component projected along the line of sight. \fig \ref{schematic} shows how the angle ($\theta$) between the sightline and the core centre increases along the line of sight. Far from the core centre the angle is small and so the velocity component along the line of sight is large. However, close to the core centre the observed velocity component is zero. Conversely, the absolute magnitude of the velocity increases towards the centre of the core. For a core with a velocity profile shallower than $v_r \propto r^{-1}$ this results in there being two positions along the line of sight with the same projected velocity. 

For an optically thin line ($\tau < 1$) all the emission directed towards the observer is visible and a Gaussian profile is observed. However, for an optically thick emission line ($\tau > 1$) only the nearest point at a given velocity is detected. In a spherical collapsing core this means that on the red-shifted side of the core, the outer layers of the core are seen, but on the blue-shifted side, the central regions are seen. For an emission line with a high critical density, there will be more emission coming from the core centre than from the outer layers of the core and consequently, the intensity is higher in the blue part of the line than the red part. The resulting line, therefore, has a central dip where the low velocity envelope obscures the core, and two peaks from the collapsing core, of which the blue peak is more intense.

The density and velocity profiles shown in the middle panel of \fig \ref{schematic} are the key requirement to produce a blue asymmetric profile. The question is, are these line of sight profiles present in filamentary collapsing cores?

\section{Method} \label{model}

\subsection{Filamentary core models}

For our models of filamentary cores we use data from a giant molecular cloud (GMC) simulation that has been the basis of much of our recent work on star formation. For details of this simulation see \citet{Smith09a}. In \citet{Smith11a} we found that `Class 0' cores from this simulation had filamentary features on scales as small as a few thousand AU. In fact, in only $\sim 25\%$ of cases could the cores truly be considered spherical. This has potentially interesting consequences for core line profiles, because, as discussed in the Introduction, previous models generally assumed spherical symmetry. 

Our original simulation used a temperature parameterisation \citep[see][]{Smith09b} which produced temperatures of around 10\,K in the dense gas ($n>10^4$ \cmc), and temperatures of up to 60\,K in the most diffuse gas ($n<10^2$ \cmc). This includes a heating term based on the YSO models of \citet{Robitaille06} that is applied in the vicinity of sink particles (which represent the locations of star formation). In order to verify that the original temperature distribution of our simulation was not the major cause of the line profile behaviour, we re-ran one of our models with a constant gas temperature of 14\,K. The only noticeable change in the result was that the central brightness temperatures were one degree higher; the qualitative behaviour of the lines did not change.

\begin{figure*}
\begin{center}
\includegraphics[width=4.9in]{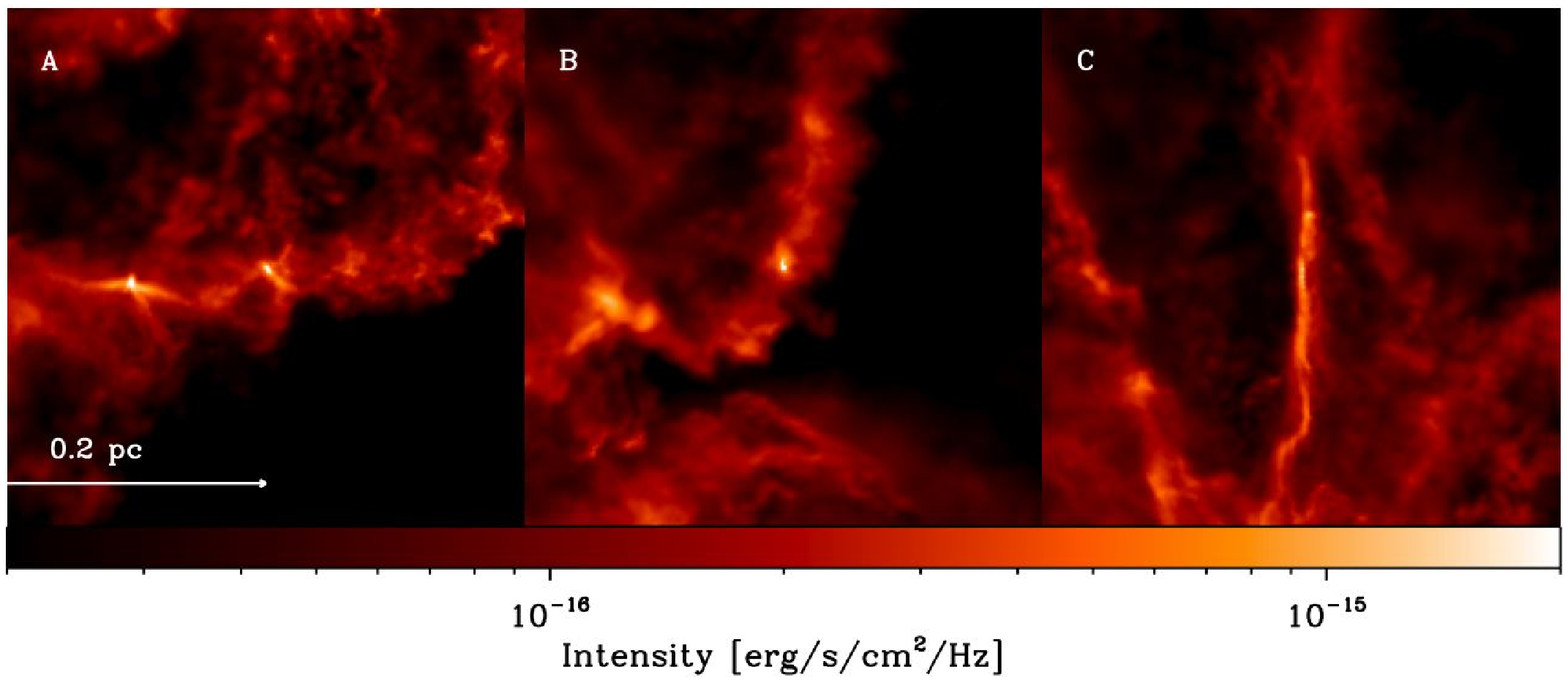}\\
\includegraphics[width=5in]{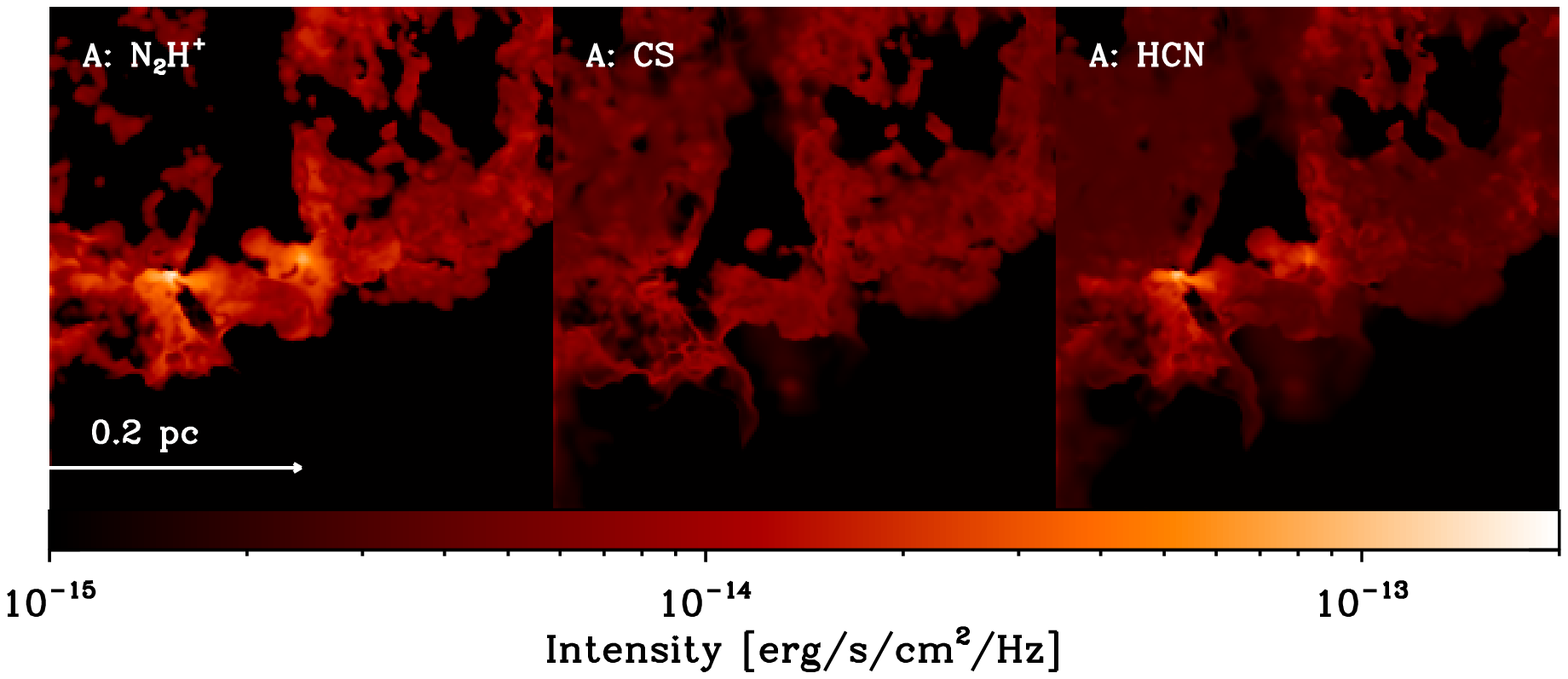}\\
\includegraphics[width=5in]{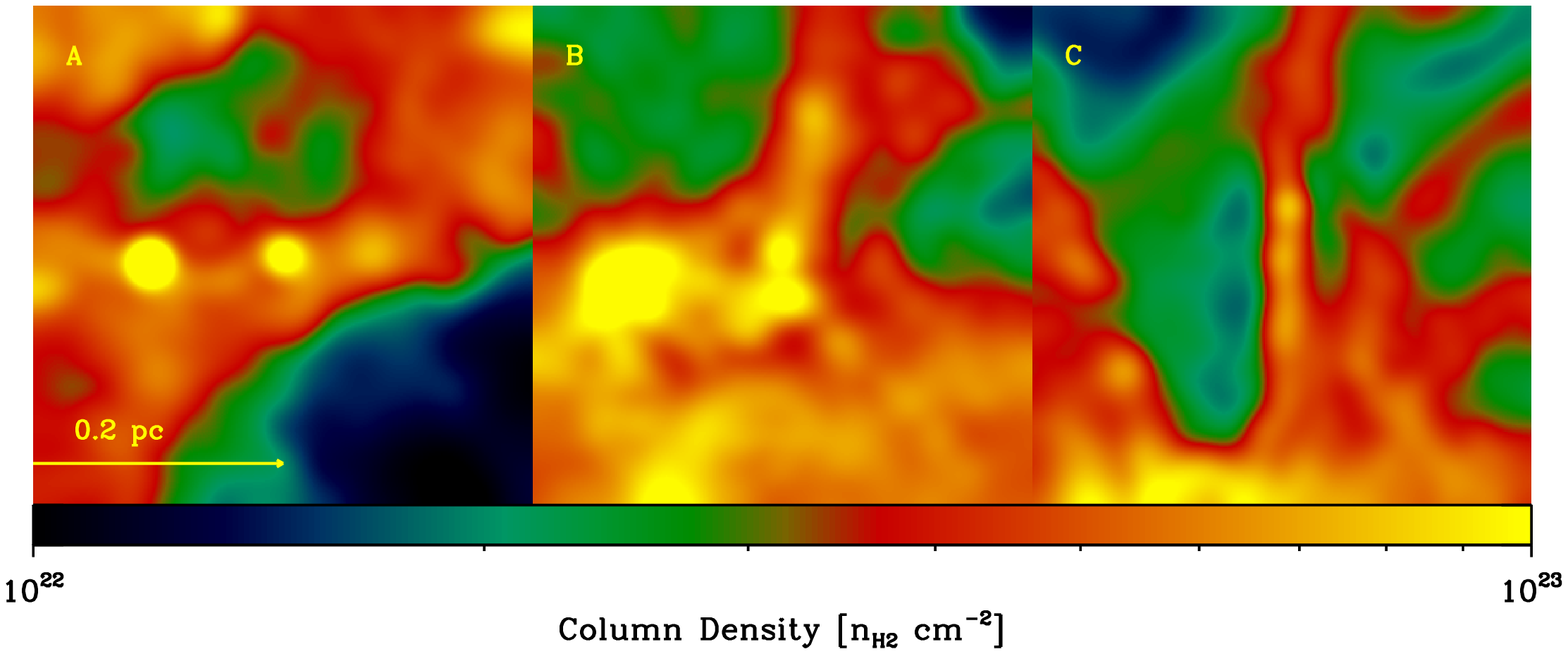}\\
\caption{\textit{Top} The three filaments in 850 \micron\ dust emission. \textit{Middle} The emission from Filament A for each species. \textit{Bottom} A column density projection of the original simulation before radiative transfer, which has been smoothed as if observed by a 0.01pc Gaussian beam}
\label{dg_image}
\end{center}
\end{figure*}

We select three typical regions from the GMC simulation each containing a core embedded within a larger filament. The cores are identified from locations of future sink-particle formation, and consequently our sample is biased as it contains only cores which truly are collapsing. We choose to select our cores in this manner so that we can 1) identify collapsing cores and 2) understand how reliably we can determine this when the core is embedded in a dense filament. The models are labeled Filaments A, B and C (for reference Filament B contains the filamentary core shown in \fig 1c of \citealt{Smith11a}). \fig \ref{dg_image} shows the 850 \micron\ dust emission calculated from each filament. Filaments A and C are aligned parallel to the axis of the model grid, and each contains a second embedded core besides the central one. Filament A is a turbulent sheet. Filament B is an interface where fast-moving diffuse gas collides with slow-moving dense gas, and has a more curved geometry. Filament C is a head-on colliding flow. These regions show marked similarities to recent \textit{Herschel} observations \citep[e.g.][]{Andre10,Menshchikov10}. We discuss the dynamics and formation mechanisms of the filaments in more detail in \sect \ref{velocities}.

In all three cases, the embedded core contains gas with densities of $3\E^4$\cmc\ to $10^7$\cmc. However when averaged across a Gaussian beam of half-width 0.01pc the maximum observed core density decreases to $10^6$\cmc, as only a small percentage of the gas in the beam has densities above this value. The filaments in which the cores are embedded have number densities of  $5\E^3$\cmc\ to $3 \E^4$\cmc. They are themselves surrounded by even more tenuous material, with densities of $10^3$\cmc\ or lower. 

We model the emission from a $0.4$ pc box centred on the collapsing core. The original SPH simulation is interpolated to a grid with a cell size of $2\E^{-3}$ pc. For a core at a distance of 150 pc this corresponds to a resolution of around 2.8 arcseconds. However, the line profiles are measured over a Gaussian beam with a full width half maximum (FWHM) of 0.01pc, which corresponds to 13.7 arcseconds. The models are selected at a time period where the central core has just started collapsing and a sink particle is about to form in the simulation. In observational parlance the cores could be considered as starless objects that are about to begin their `Class 0' phase of core evolution. 

In order to make a comparison to observational data we estimate the mass of the filaments by calculating the total mass of gas in our grid models with a density in the range $5\E^3$\cmc\ $< n < 3 \E^4$\cmc. For Filaments A, B and C we obtain values of 6.4 \msun, 12.3 \msun and 14.7 \msun respectively. \citet{Hacar11} find values in the range 4.8 \msun to 11.3 \msun for four filaments they studied in L1517. Our values are therefore in good agreement with the masses of observed filaments.

Furthermore, we also find good agreement when comparing our simulation to observed column densities in star-forming regions. The bottom panel of \fig \ref{dg_image} shows a column density projection from the original simulation of each filament before radiative transfer, which has been smoothed as if observed by a 0.01pc width Gaussian beam. The column densities of our filaments are in the range $10^{22} -10^{23}$ \cmsq\ with a typical column density of around $4\E^{22}$ \cmsq, and an increase in column density at the location of the dense embedded cores. \citet{Menshchikov10} find column densities in the range $5\E^{20}-1.4\E^{23}$ \cmsq\ for filaments in Aquila, and column densities of $3\E^{20}-8.6\E^{21}$ \cmsq\ for filaments in Polaris. Our filaments resemble dense filaments in Aquila far more than the Polaris filaments. This is reasonable as they are extracted from a simulation undergoing rapid star formation like that seen in the Aquila field. The Polaris region contains no clear examples of pre-stellar cores \citep{Ward-Thompson10}. In conclusion our model filaments closest observational analogue is to dense star-forming filaments in clustered regions rather than diffuse quiescent filaments.

\subsection{Molecular line modeling}
We use the radiative transfer code RADMC-3D \footnote{This code is publicly available with the permission of the author, Cornelius Dullemond, at the website http://www.ita.uni-heidelberg.de/dullemond/software/radmc-3d/. A paper describing its usage is currently in preparation.} to carry out the line modelling of the cores. RADMC-3D models a variety of radiative processes including dust thermal emission and absorption, dust scattering, as well as gas atomic and molecular lines. It is this latter capability which is of importance here.

The chemical species which we will consider in this paper have emission lines with high critical densities and as such are typically not in local thermal equilibrium (LTE). Consequently, a non-LTE approach must be used for the line transfer. We use the large velocity gradient (LVG) approximation \citep{Sobolev57} which uses local velocity gradients to define photon escape probabilities. For a full description of how LVG has been implemented in RADMC-3D see \citet[][a]{Shetty11}. We also use a method known as `Doppler catching' to interpolate under-resolved velocities along each line of sight as described in \citet[][b]{Shetty11b}

In the cores studied here a given line of sight velocity may occur at more than one position and so the assumption of LVG is not fully satisfied. However, \citet{Ossenkopf97} showed that the Sobolev approximation can be used in regimes where it does not strictly apply, such as turbulent flows in molecular clouds, and still yield reasonably accurate results. In spherical homogeneous flows like the idealised case shown in \fig \ref{theory}, he found that the error had a maximum value of only $20\%$. Moreover this error was mainly in the line intensity, rather than in the shape of the line profile. We test this further by calculating the LTE line profiles for our cores and then comparing them to the LVG results. In LTE the maximum intensities are higher due to the artificially high level populations. However, both methods show a dip at the same velocity in the line profile, and so we can be confident that our assumption of LVG is not affecting the qualitative behaviour of the lines. We also tested our method using a completely spherical collapsing core for an optically thick species and confirmed that the expected blue asymmetric line profile was produced in agreement with previous studies.
\subsection{Molecular Species}

The general procedure for creating the molecular line profiles is the following. The filaments from our original simulation are mapped onto a grid and the number density of each desired species is calculated using simplified prescriptions for the abundances as a function of density. The grids are used as inputs for RADMC-3D, which calculates the level populations of the tracer species, and then carries out ray-tracing to calculate the molecular emission. To compute the level populations of our chemical species we use line data obtained from the Leiden LAMDA database \citep{Schoier05} and assume that molecular hydrogen is the dominant collisional partner.

We consider three molecular species in this study, each having different properties and tracing different regimes, as shown in \tab \ref{species}. The chosen line transitions all lie within the observable range of most ground based detectors, such as ALMA. All transitions have a high critical density, and hence are dense gas tracers. Consequently, any differences in the emitted line profiles are primarily due to optical depth effects.

\begin{table*}
	\centering
	\caption{The properties of the molecular tracers. The critical density for LTE is estimated using the relation $n_{H_2}=A_{ul}/K_{ul}$ where $A_{ul}$ is the Einstein A coefficient and $K_{ul}$ is the collisional rate coefficient at an assumed kinetic temperature of 20K. }
		\begin{tabular}{l l l l l }
   	         \hline
	         \hline
	         Tacer & Line & Critical density & Optically & Abundance Law\\
	         & & [cm$^{-3}$]& & [n/n$_{H_2}$]\\
	         \hline
	         \n  & 1-0 & 1.4 $\E^5$ & thin & $10^{-10}$ \\
	         CS  & 2-1 & 3.2 $\E^5$  & thick & $4\E^{-9} e^{-n_{H2}/n_d}$  \small{($n_d=4\E^4$ \cmc)}\\
	         HCN  & 1-0 & 2.6 $\E^6$& thick & $3\E^{-9}$ \\
		\hline
		\end{tabular}
	\label{species}
\end{table*}

The isolated component of \n\ is optically thin and traces the entirety of the dense gas. Molecular collisional rates and the the Einstein coefficients of this species have been calculated by \citet{Daniel05} and \citet{Schoier05}. \n\ has seven hyperfine lines that make up its (1-0) transition. We model the isolated 101-012 hyperfine line, which has a rest frequency of 93.176 GHz. The critical density for LTE is estimated to be $n_{crit}=1.4\E^5$ cm$^{-3}$, using the relation $n_{crit}=A_{ul}/K_{ul}$ where $A_{ul}$ is the Einstein A coefficient and $K_{ul}$ is the collisional rate (at an assumed kinetic temperature of 20K). As the isolated hyperfine line of \n\ is optically thin it contains information about all the velocities along the line of sight and no asymmetric infall signature is expected. Nonetheless, this line is a useful reference when compared to optically thick lines to separate self-absorption effects from intrinsic variation in the velocity profiles. 

We calculate the abundance of \n\ using the results of \citet{Aikawa05} who find an abundance of $A_{\mathrm{N2H}^+}=10^{-10}$ relative to the H$_2$ number density of the gas. The abundance of \n\ has been proposed to rise when CO is frozen onto dust grains as CO destroys \n. For instance \citet{Jorgensen04} found that as CO becomes depleted onto dust grains the abundance of \n\ increases to $A=10^{-8}$. However, when we used this value for the \n\ abundance above the CO freeze out density of $n=3\E^4$ cm$^{-3}$ \citep{Bergin07}, we found that the \n\ emission became too bright and optically thick compared to its observed behaviour. In \citet{Aikawa05} the \n\ abundances in collapsing Bonnor-Ebert spheres showed variations of at most a factor of a few, and a slight decrease in abundance at the core centre. Therefore we will use the simple constant abundance model and leave a more detailed analysis of the \n\ chemistry to future works.

The second molecular tracer is CS, which is optically thick and traces the more diffuse gas in the envelope around the core. Molecular collisional rates and Einstein coefficients for CS were calculated by \citet{Turner92}. We model the 2-1 transition of CS, which has a rest frequency of 98.981 GHz. The critical density of CS is $n_{crit}=3.2\E^5$ cm$^{-3}$ at a collisional temperature of 20K. As an optically thick line, CS exhibits self-absorption and is capable of producing a blue asymmetric collapse profile. The classic work on infall signatures by \citet{Zhou92} used CS. However, \citet{Tafalla02} showed that CS is depleted at the centre of cores due to its freezing onto dust grains. They found the depletion could be described by an abundance prescription
\begin{equation}
A_{\mathrm{CS}}(r)=A_0 e^{-n(r)/n_d}
\end{equation}
 where $A_{\mathrm{CS}}$ is the abundance, and $n$ is the H$_2$ number density. We adopt values of $n_d=2\E^{4}$\cmc\ for the depletion density and $A_0=4\E^{-9}$ for the low density abundance limit, based on the typical values given in \citet{Tafalla02}. \citet{Stahler10} recently carried out line modelling of diffuse cores with and without CS depletion, and found that the models with depletion were a better match to observations, and so we adopt it here. Due to depletion the number density of CS is almost negligible in the densest regions of our cores.

The third molecular species we consider is HCN, which is optically thick and traces the dense gas. The collisional rates for this species have been calculated by \citet{Green74} and \citet{Dumouchel10}. HCN is particularly useful because it has three well-separated hyperfine emission lines within its 1-0 level and each one has slightly different optical depths, with F(0-1) being the optically thinnest, F(2-1) the optically thickest, and F(1-1) intermediate in optical depth. \citet{Sohn07} have recently pointed out the utility of this species as a diagnostic tracer of infall motions in dense cores and as such it is the key tracer we consider here for our filamentary cores. We use a constant abundance of $A_{\mathrm{HCN}}=3\E^{-9}$ for the molecular abundance of HCN, following \citet{LeeJ04}. Note that some variation in abundance is seen in core centres by \citet{LeeJ04} but for simplicity we neglect this here. The rest frequency of the central HCN hyperfine line is 88.632 GHz.

To check that the cores are visible through the filament, we first calculate the dust emission in RADMC-3D, using a dust-to-gas ratio of 0.01 and a constant dust temperature of 20\,K. \fig \ref{dg_image} shows that the embedded cores can be clearly seen in 850 \micron\ dust emission.

The line models are calculated for each molecular species at 256 evenly-spaced wavelengths, centred on the rest wavelength of the relevant emission line. The bottom panels of  \fig \ref{dg_image} show the emission from Filament A at the rest wavelength for each of the three species (for HCN we show the F(2-1) hyperfine line). The \n\ emission is mainly from the dense gas traced by the dust emission, whereas the CS shows more extended emission from the surrounding filament. The HCN emission is bright in both the cores and the surrounding regions. The second core, to the left of the central core, actually appears brighter since it is more evolved and has a higher temperature.

\begin{figure}
\begin{center}
\includegraphics[width=3.5in]{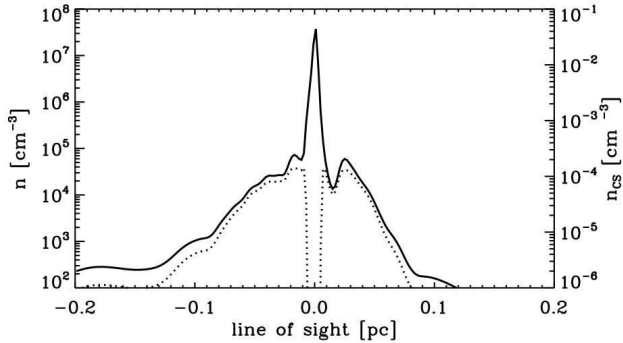}
\caption{The number density of CS \textit{(dotted line)} and the gas number density \textit{(solid line)} measured along a line through the core centre. There is a sharp rise in density at the location of the collapsing core (at a number density of around $3\E^4$ \cmc). However at these densities the CS is frozen out and so it only traces the surrounding filament.}
\label{freeze_out}
\end{center}
\end{figure}

The abundance of CS is lower in the core centre than in the surrounding filament due to the effects of freeze out (outlined in the previous Section). \fig \ref{freeze_out} shows the number density of CS and the gas density along a line through the core centre. The densest regions of the core are effectively not contributing to the CS emission. The HCN and \n, however, follow the gas distribution. The gas densities are extremely low outside the filament, but increases steadily in the filament, and rises sharply at the center of the embedded core. As the density of the gas is so low at the edges of our box, we can be confident that we are integrating over the full length scale from which there are contributions to the total emission.

\section{Results}\label{results}
\subsection{Line Profiles}\label{lines}

\begin{figure*}
\begin{center}
\includegraphics[width=5.5in]{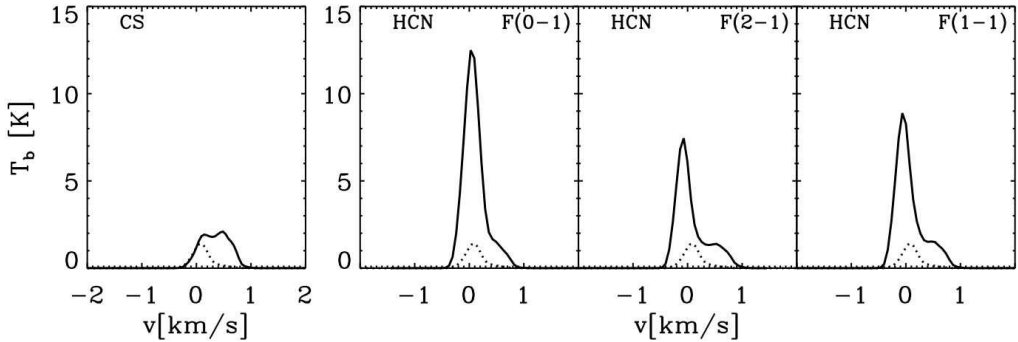}\\
\caption{The emission line profiles from the embedded core at the centre of Filament A calculated using a simulated 0.01pc half-width beam through the centre of the core. The dotted lines show the \n (1-0) optically thin line emission from the core. Higher intensity emission is seen from the blue side of the HCN F(2-1) line profile when the model is seen from this viewing angle.}
\label{lines_0}
\end{center}
\end{figure*}

In \fig \ref{lines_0} we show the line profiles of the central core of Filament A, calculated for a beam with a FWHM of 0.01pc. The beam passes directly through the centre of the core and is viewed at a model inclination and rotation of zero. The dotted lines show the line profile of \n. As \n \ is optically thin, all the emission is visible and the line profiles are generally Gaussian. This confirms that deviations from this profile in the other lines are due to optical depth effects. 

In \fig \ref{lines_0} the CS profile shows a slight red asymmetry rather than a blue one as the blue emission has been largely self-absorbed. However, higher intensity emission is seen from the blue side of the HCN F(2-1) line profile. The thinner F(0-1) line shows only a shoulder on the red side of a central peak, whereas the F(2-1) hyperfine line gives a clearer blue asymmetric profile. 

Generally the lines have more emission in their wings when observed in CS and HCN than in \n\ emission. The bottom panel of  \fig \ref{dg_image} shows the emission in each species from this filament. Due to their higher abundances, the emission from CS and HCN extends beyond the densest parts of the filament. This greater contribution of material from the outer edges of the filament increases the apparent line width of the core. The filament velocities will be discussed in more detail in \sect \ref{linewidths}.

\begin{figure*}
\begin{center}
\begin{tabular}{c c}
\includegraphics[width=3in]{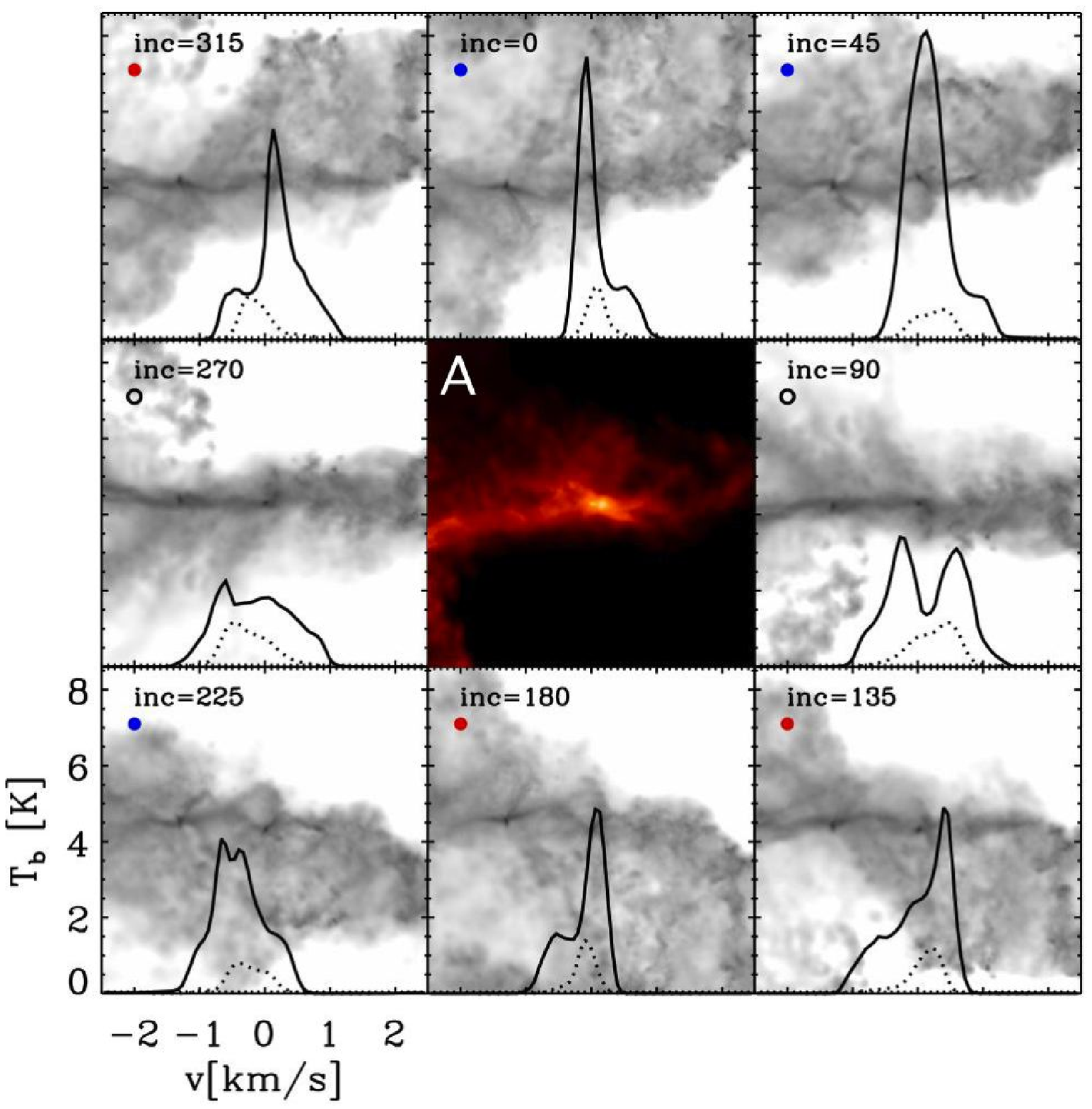}
\includegraphics[width=3in]{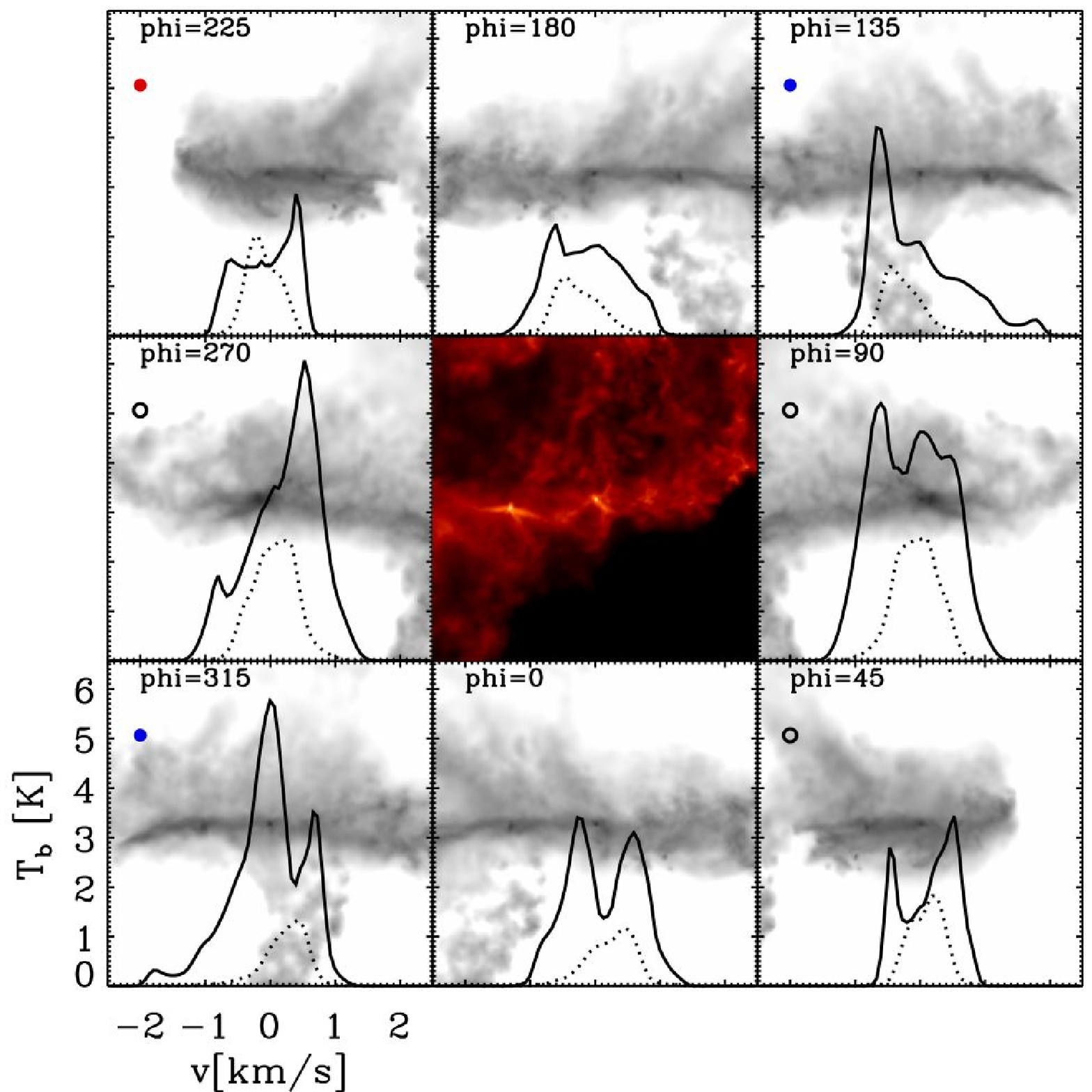}\\
\includegraphics[width=3in]{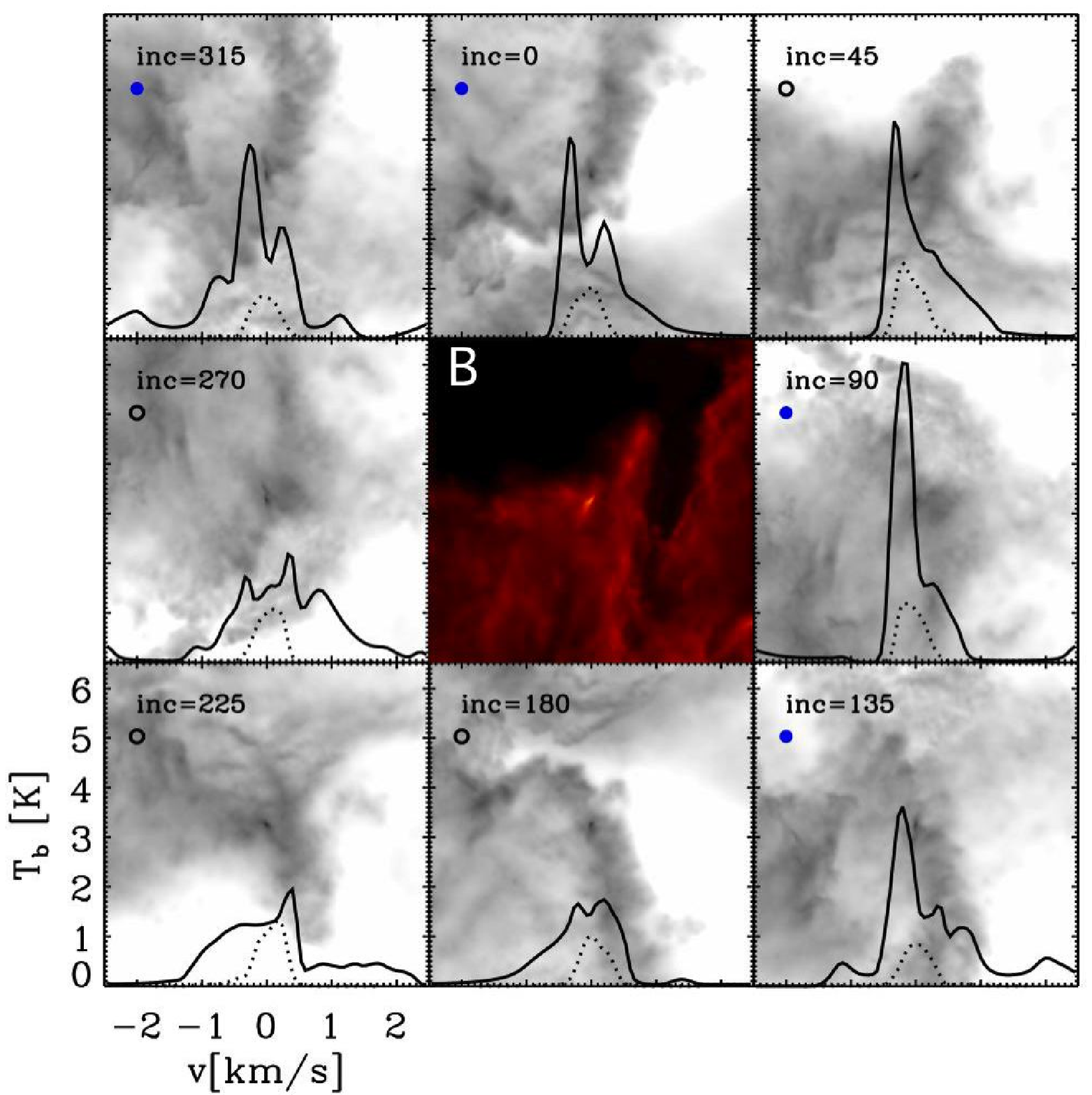}
\includegraphics[width=3in]{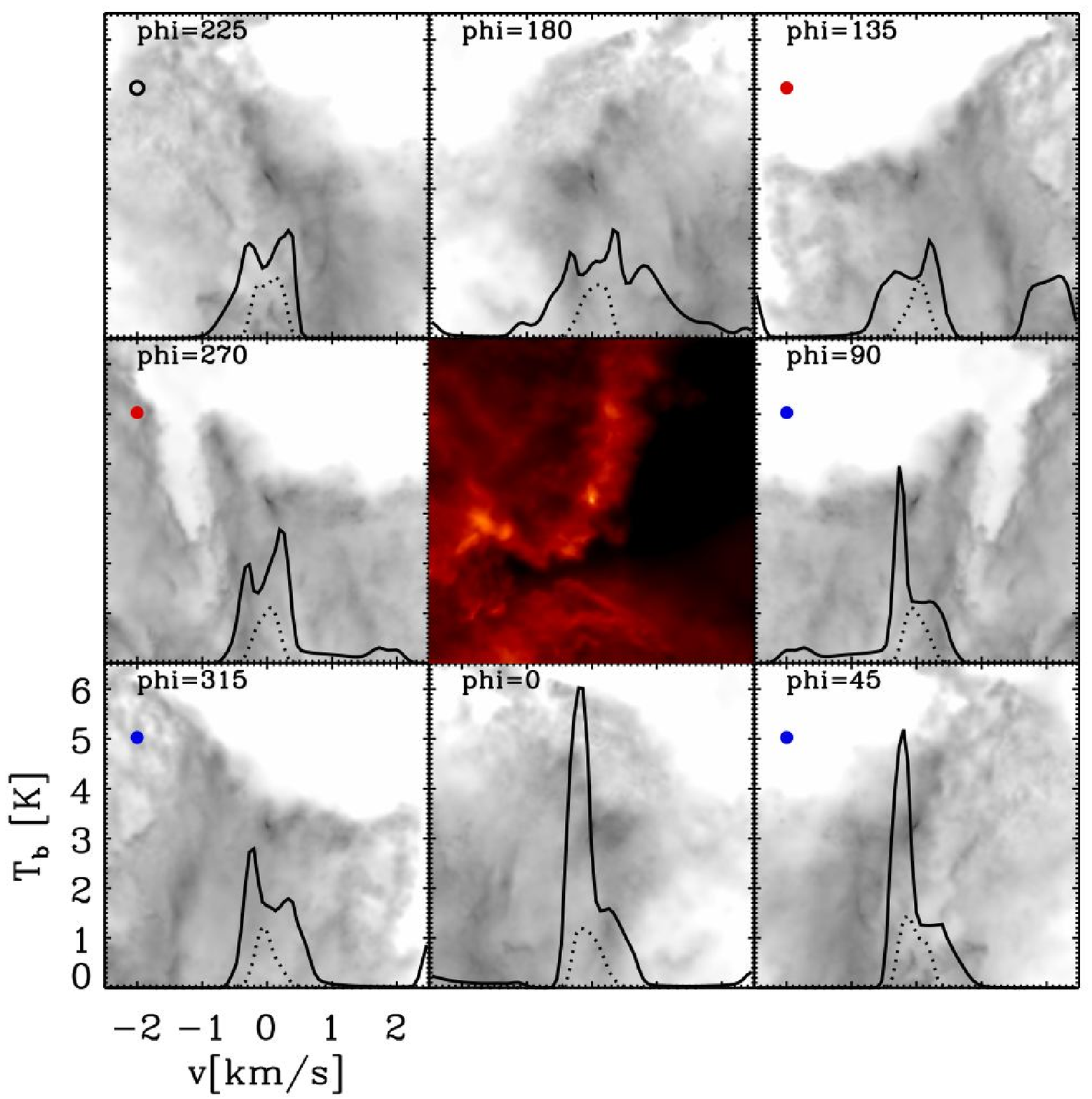}\\
\end{tabular}
\caption{The HCN F(2-1) line profiles of Filaments A \textit{(top)}, B \textit{(bottom)} at different viewing angles. The central colour image of each panel shows the dust density in the plane in which the sight-lines pass through the core. The background grey-scale images show the large scale (0.4 pc diameter) dust emission map of the filament when viewed at the labeled angle. The position at which the outer images touch the central image denotes the orientation of the sightline. The line profiles of the central core at each angle are over-plotted on top of the grey scale image. A coloured dot shows how the profile was classified. The line profiles are calculated for a 0.01pc beam centred directly on the embedded core. In the left hand panels the rotation angle has a constant value of $\phi=0$\degree, and in the right panels the inclination has a constant value of inc=90\degree}
\label{sightlines}
\end{center}
\end{figure*}

\begin{figure*}
\begin{center}
\begin{tabular}{c c}
\includegraphics[width=3in]{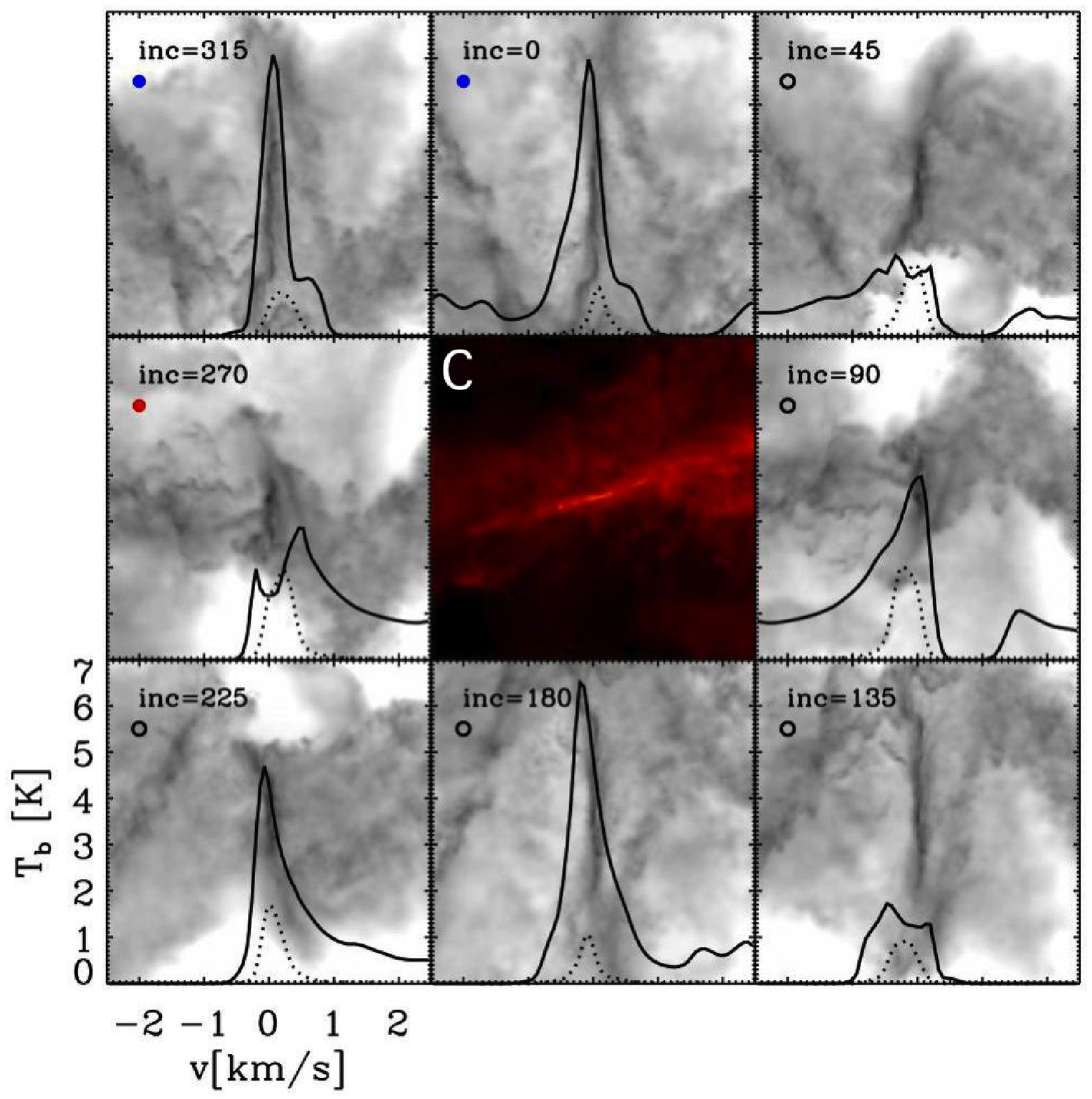}
\includegraphics[width=3in]{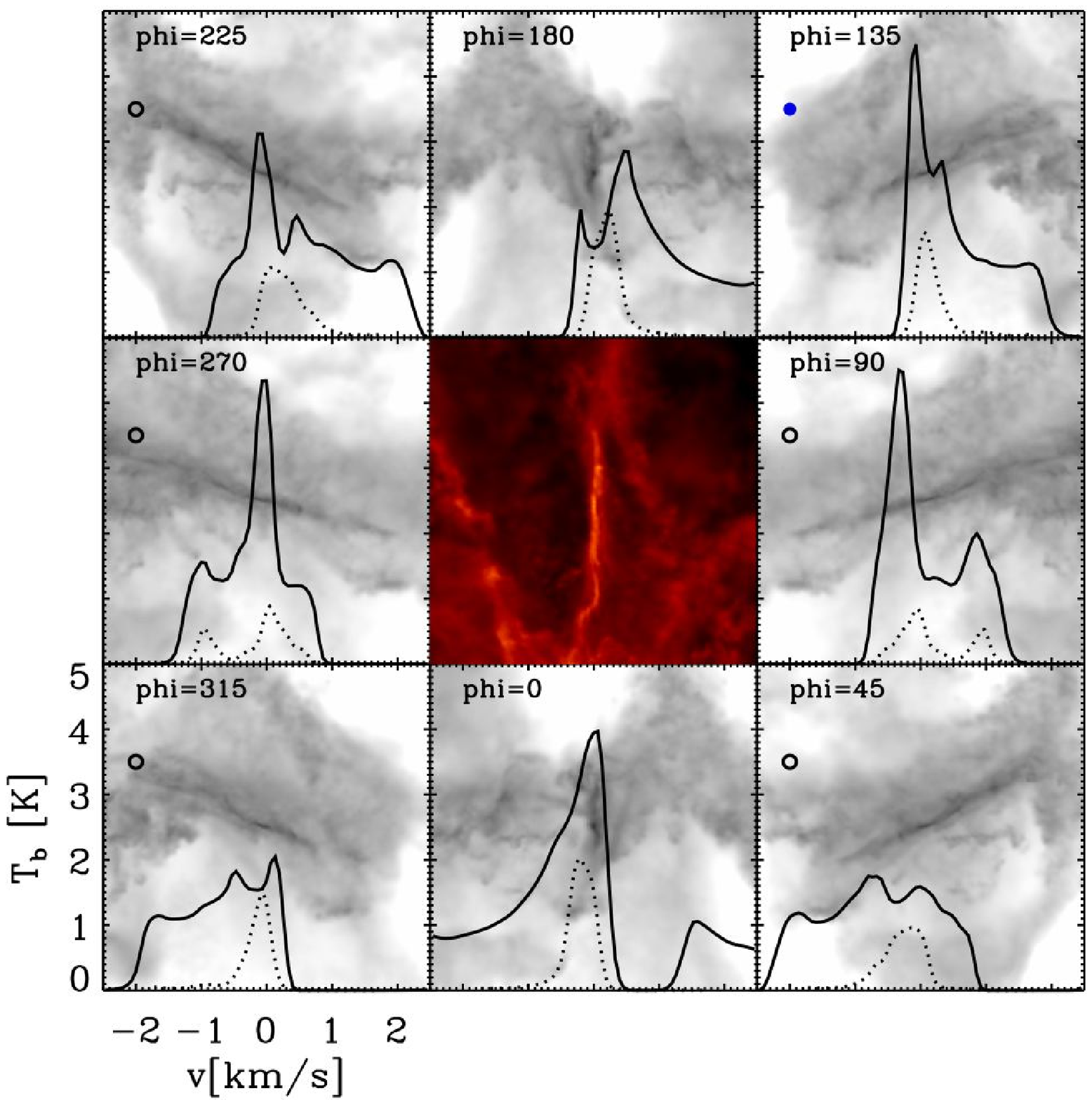}\\
\end{tabular}
\caption{The HCN F(2-1) line profiles of Filament C at different viewing angles. As in \fig \ref{sightlines}.}
\label{sightlinesC}
\end{center}
\end{figure*}

\begin{figure*}
\begin{center}
\begin{tabular}{c c}
\includegraphics[width=3.1in]{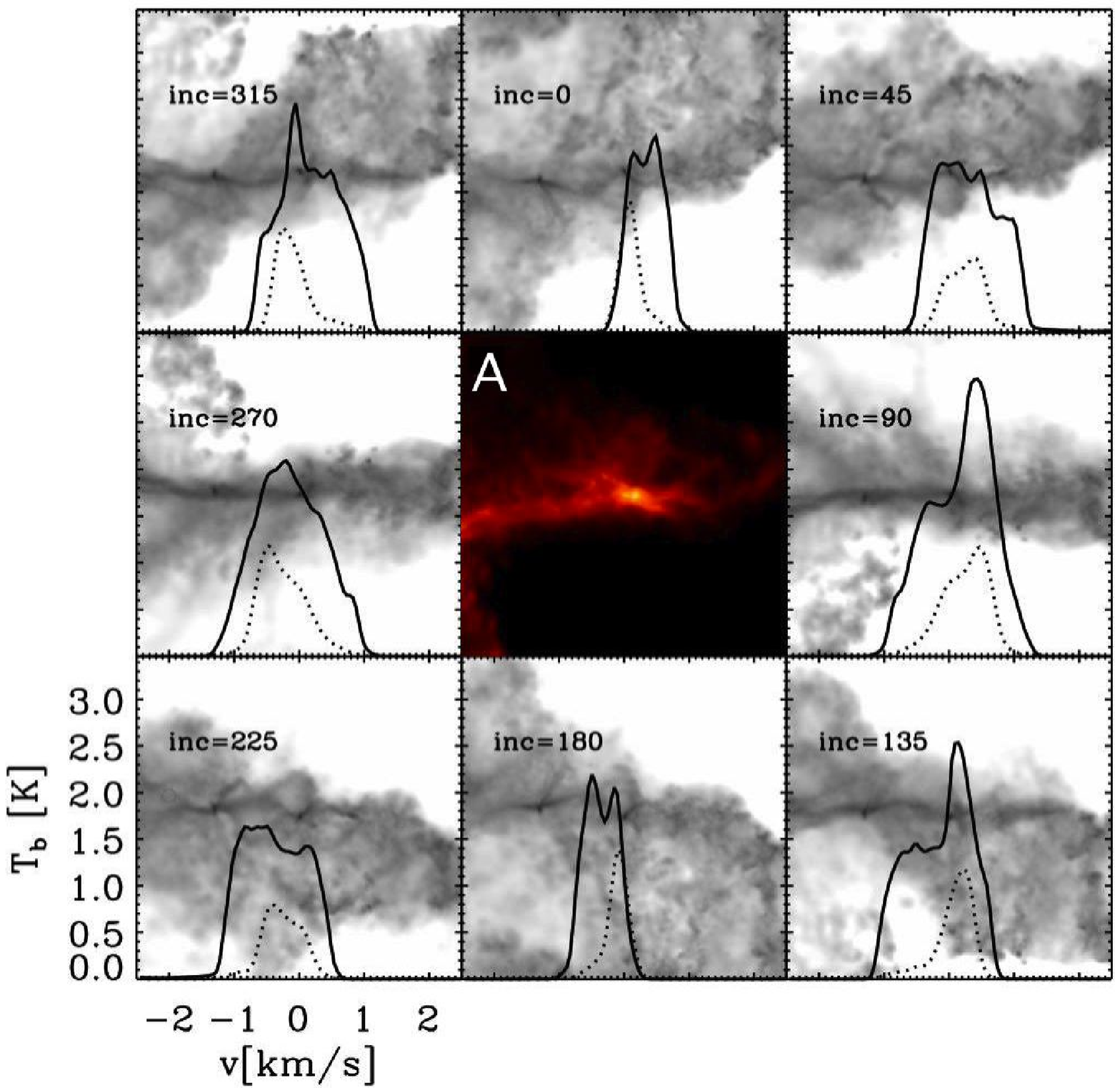}
\includegraphics[width=3in]{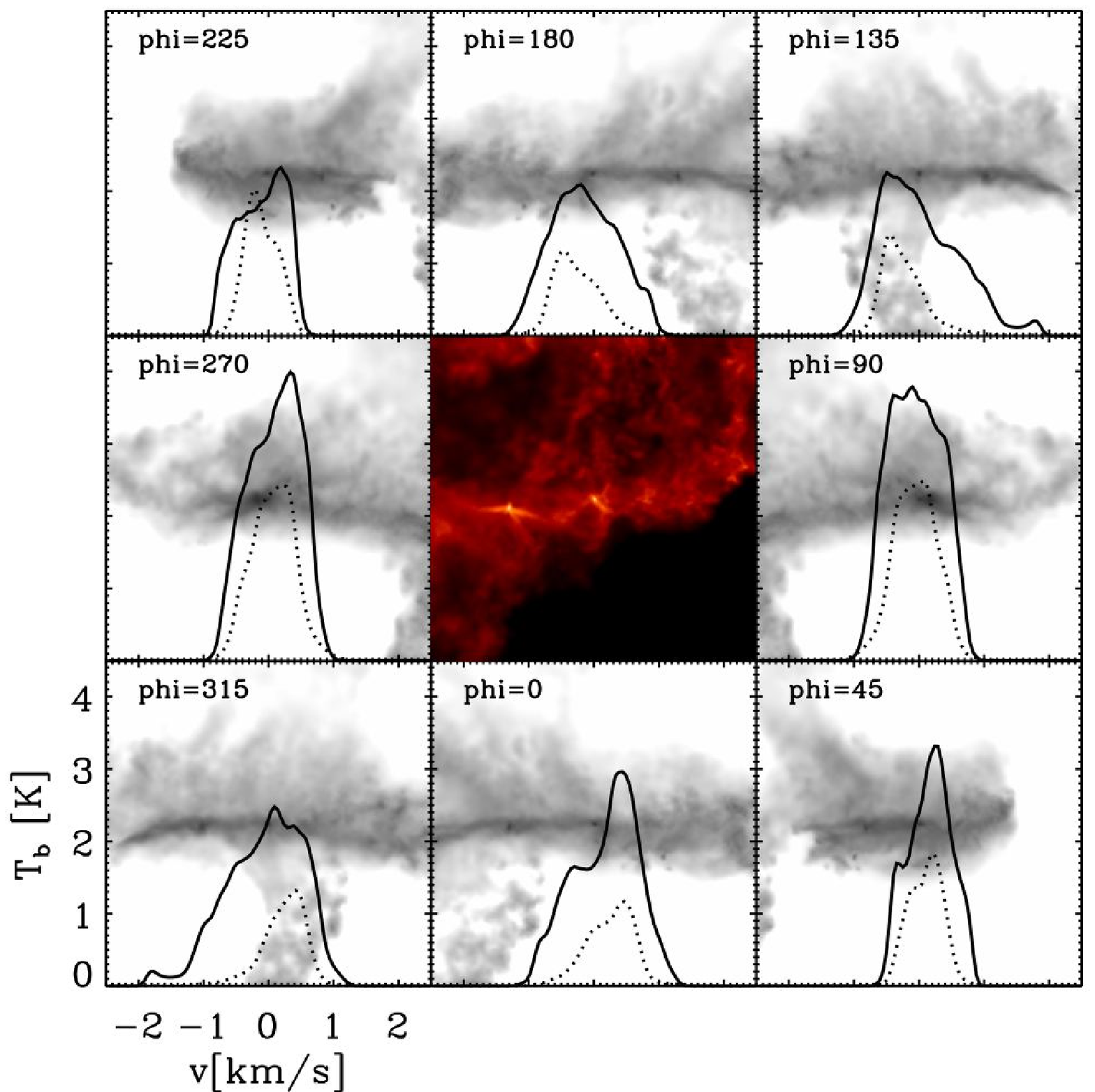}\\
\includegraphics[width=3in]{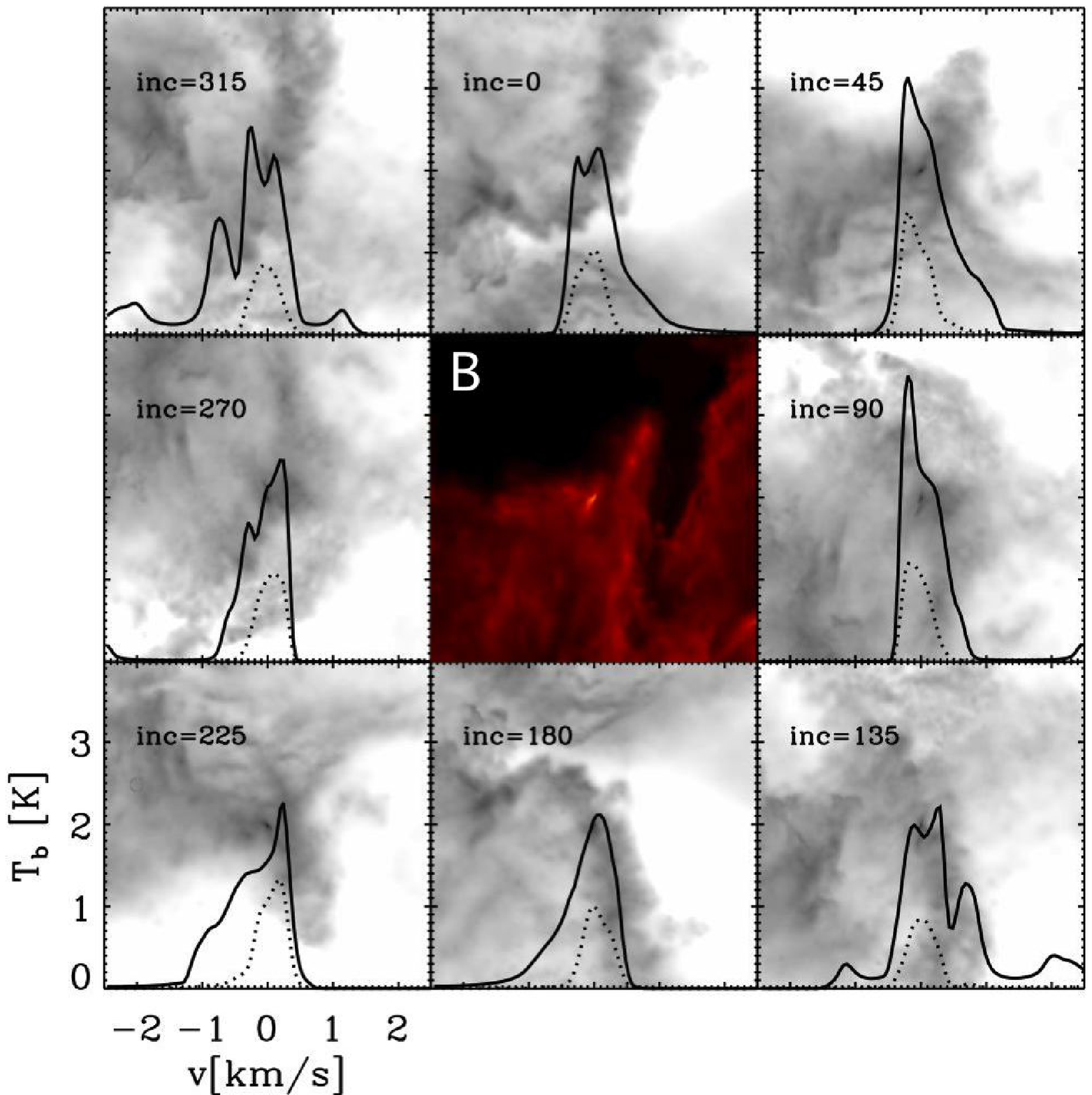}
\includegraphics[width=3in]{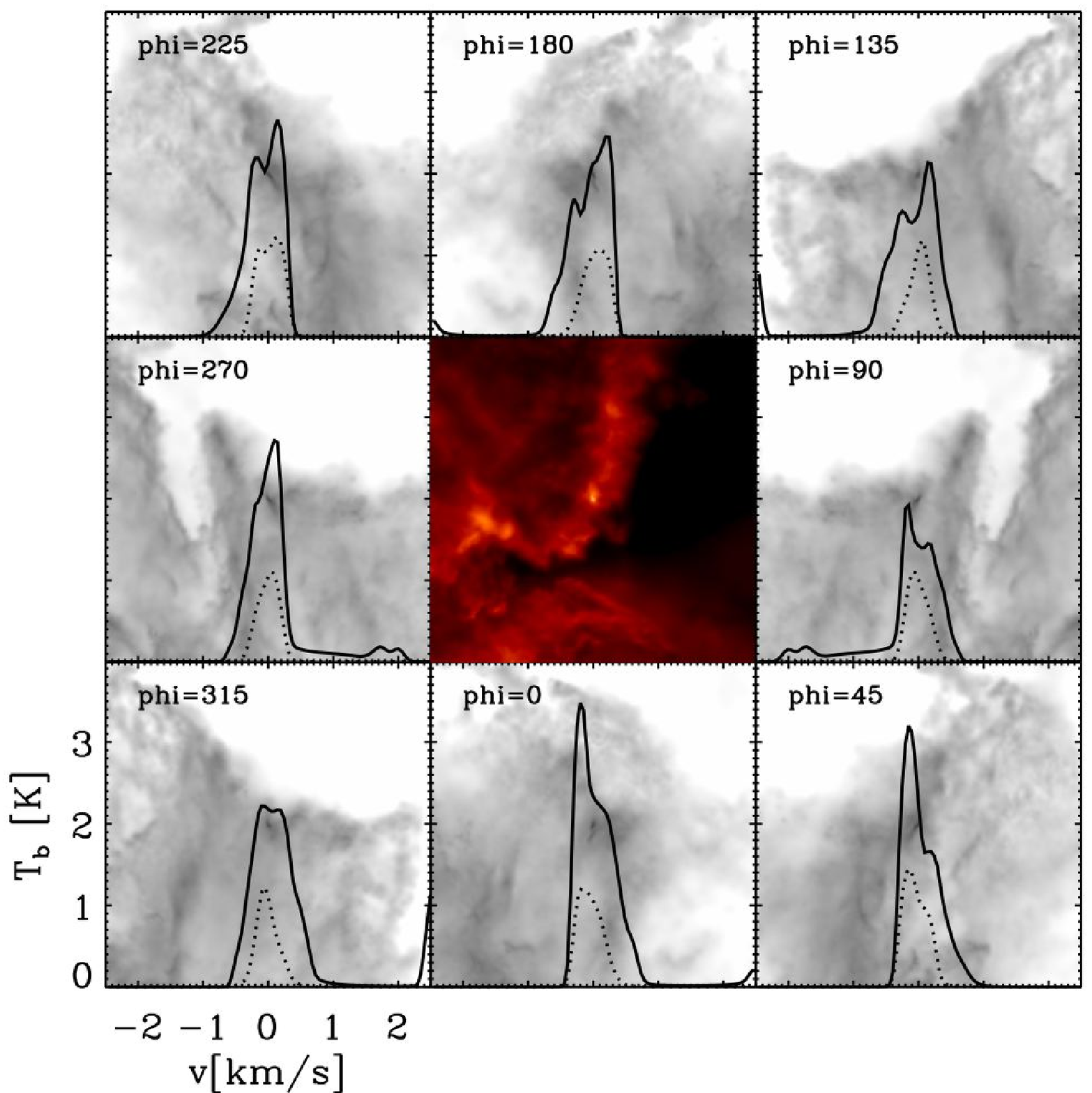}\\
\end{tabular}
\caption{As in \fig \ref{sightlines} but for the case of CS(2-1) emission.}
\label{sightlines_cs}
\end{center}
\end{figure*}

\begin{figure*}
\begin{center}
\begin{tabular}{c c}
\includegraphics[width=3in]{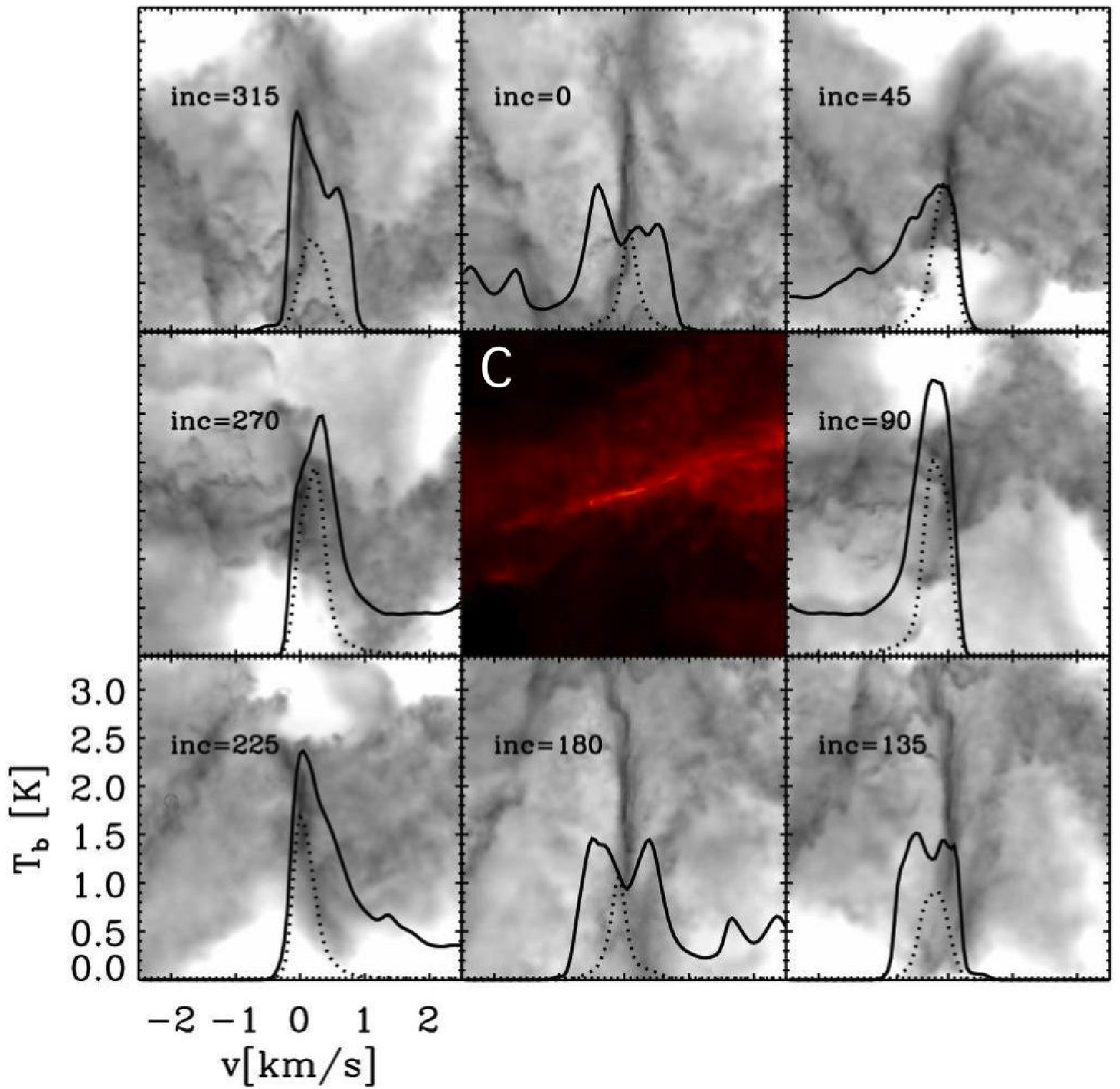}
\includegraphics[width=3in]{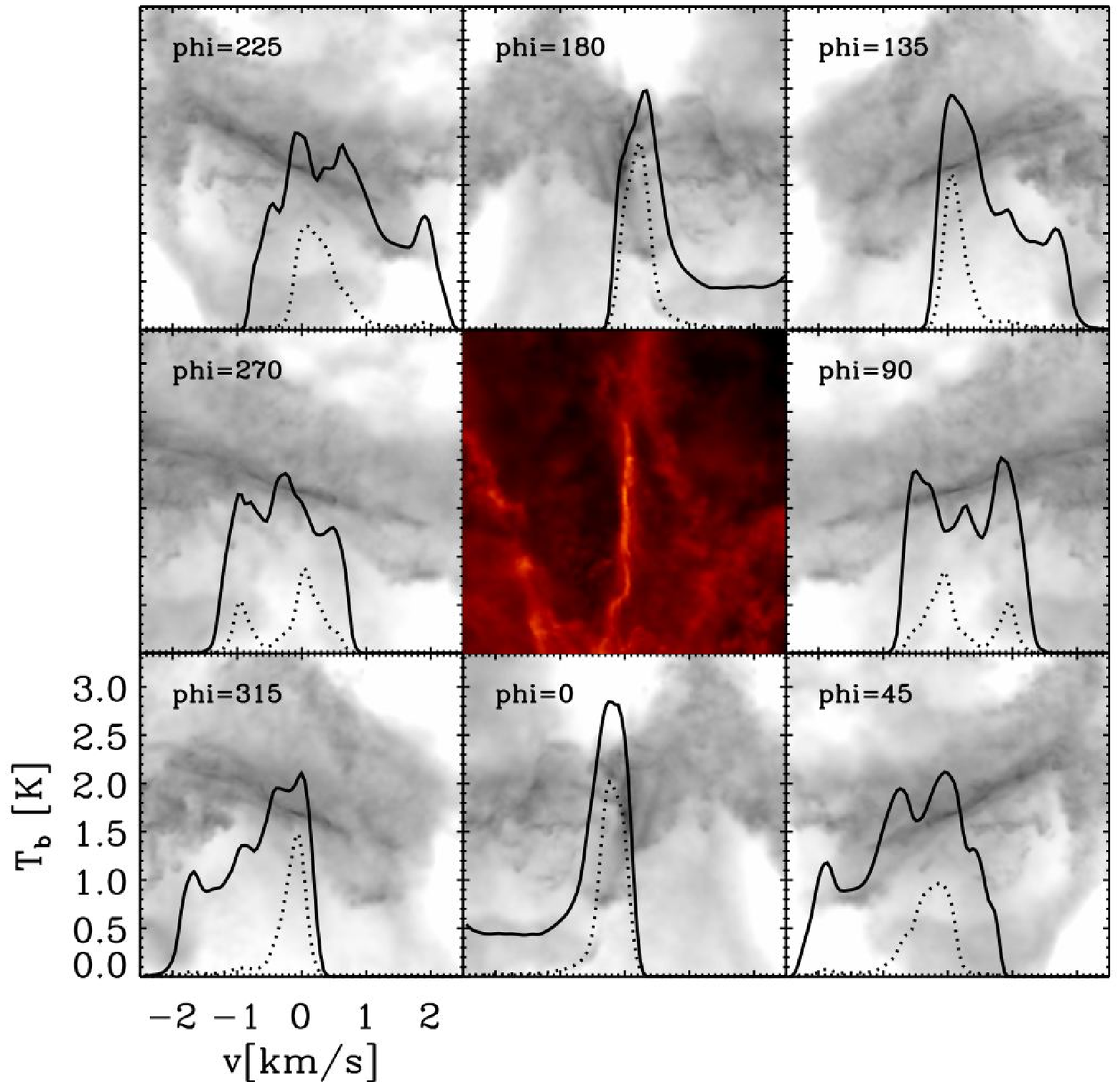}\\
\end{tabular}
\caption{As in \fig \ref{sightlines} but for the case of CS(2-1) emission.}
\label{sightlinesC_cs}
\end{center}
\end{figure*}

\fig \ref{lines_0} shows a single viewing angle corresponding to an inclination and rotation of zero in our co-ordinate system. However, in such a filamentary and complex environment, the gas properties are likely to vary along each sightline. This is examined in Figures \ref{sightlines} and \ref{sightlinesC} which shows the HCN F(2-1) hyperfine line at various viewing angles around the filaments; the CS(2-1) lines are shown in Figures \ref{sightlines_cs} and \ref{sightlinesC_cs}. In the left hand panels of the figures the model is kept a the default rotation, $\phi$=0\degree, and we view the model at 45\degree\ intervals in inclination. In the right hand panels, we keep a constant inclination of inc=90\degree\ and view the model at 45\degree\ intervals in rotation. We sample a total of 14 unique lines of sight through each core centre.

It is immediately apparent that there is a large degree of variability among the core line profiles depending on the viewing angle. Importantly, this is also true of sight-lines which pass through the same material but in the opposite direction. For example, the HCN lines at inc=0\degree\ and inc=180\degree\ in Filament A pass through the same gaseous regions but in the opposite direction, yet the former has a blue asymmetric profile and the latter a red. Even more striking examples in Filament A are the phi=90\degree\ and inc=270\degree\ sight-lines which do not even share the same general shape. This variability in the line profiles has immediate unfortunate implications for the interpretation of the optically thick line profiles for cores embedded within filaments. If a simple dynamical model (for instance that of a collapsing Bonnor-Ebert sphere) were used to predict observational line profiles, very different interpretations of the dynamical state of this core would be deduced, purely due to line-of-sight effects. It is worth stating again that the core in the centre of each filament is the same regardless of viewing angle.

\fig \ref{sightlines} contains sight-lines which show the expected blue asymmetry, but the majority of cases have no clear asymmetry at all. There are even a few cases that display a red asymmetry. A red profile is commonly taken as a signature of core expansion, or oscillation \cite[e.g][]{Keto06} as the reversed velocity field means that the red portion of the line profile originates from the higher density part of a spherical core. \tab \ref{nblue} shows the total number of visible blue and red asymmetries in the HCN F(2-1) for our three embedded cores. We classify the cores as blue when there was a clear peak in the blue side of the profile and a clear lower intensity peak or shoulder in the red. When the more intense peak was on the red side of the profile, we classify the core as red. If both peaks were of a similar intensity, two clear peaks could not be distinguished within the line, or there was only one clear peak, the line profile is classified as being ambiguous.

\begin{table}
	\centering
	\caption{A summary of the classification types assigned to the HCN F(2-1) lines from filaments A, B and C using the line profile shapes. Despite the fact that the embedded cores are collapsing, a blue asymmetric line profile is seen in only 36\% of cases. }
		\begin{tabular}{l l c c c}
   	         \hline
	         \hline
	         Filament & Blue & Red & Ambiguous \\
	         \hline
	         A   & 5/14 & 4/14 & 5/14 \\
	         B   &7/14 & 2/14 & 5/14 \\
	         C   &3/14 & 1/14 & 10/14\\
	         
	         Total & 15/42 & 7/42 &20/42 \\
		 \hline	         
	          & 36\% & 17\% &47\% \\
		\hline
		\end{tabular}
	\label{nblue}
\end{table}

For HCN under half of the line profiles are blue, with a total of 35\% of the sight-lines showing this asymmetry. Of the remainder, 17\% of sight-lines show a red asymmetry, and 48\% of cases show no clear asymmetry. For CS the interpretation is even more complicated as the resulting line profile is mainly determined by the dynamics of the filamentary envelope surrounding the core (since molecular freeze out of CS onto dust grains leads to very low abundances at the core centre, as shown in \fig \ref{freeze_out}). \fig \ref{sightlines_cs} shows that the CS line profiles are as likely to show a red asymmetric profile as a blue asymmetry and that the majority of the lines are hard to interpret, particularly in filaments A and C. \citet{Park98} showed that underlying `clumpiness' within a core can flatten the resulting line profiles and superpose small velocity features onto the line where the maximum clump velocities lie. This complicates the interpretation of the line profile in irregular density fields, such as those in our models, and in some extreme cases can even lead to false infall expansion signatures. In our models we find only six line profiles which clearly resemble classical infall or expansion profiles in the CS sample, three (7\%) red, and three (7\%) blue. 

\begin{figure}
\begin{center}
\includegraphics[width=2.9in]{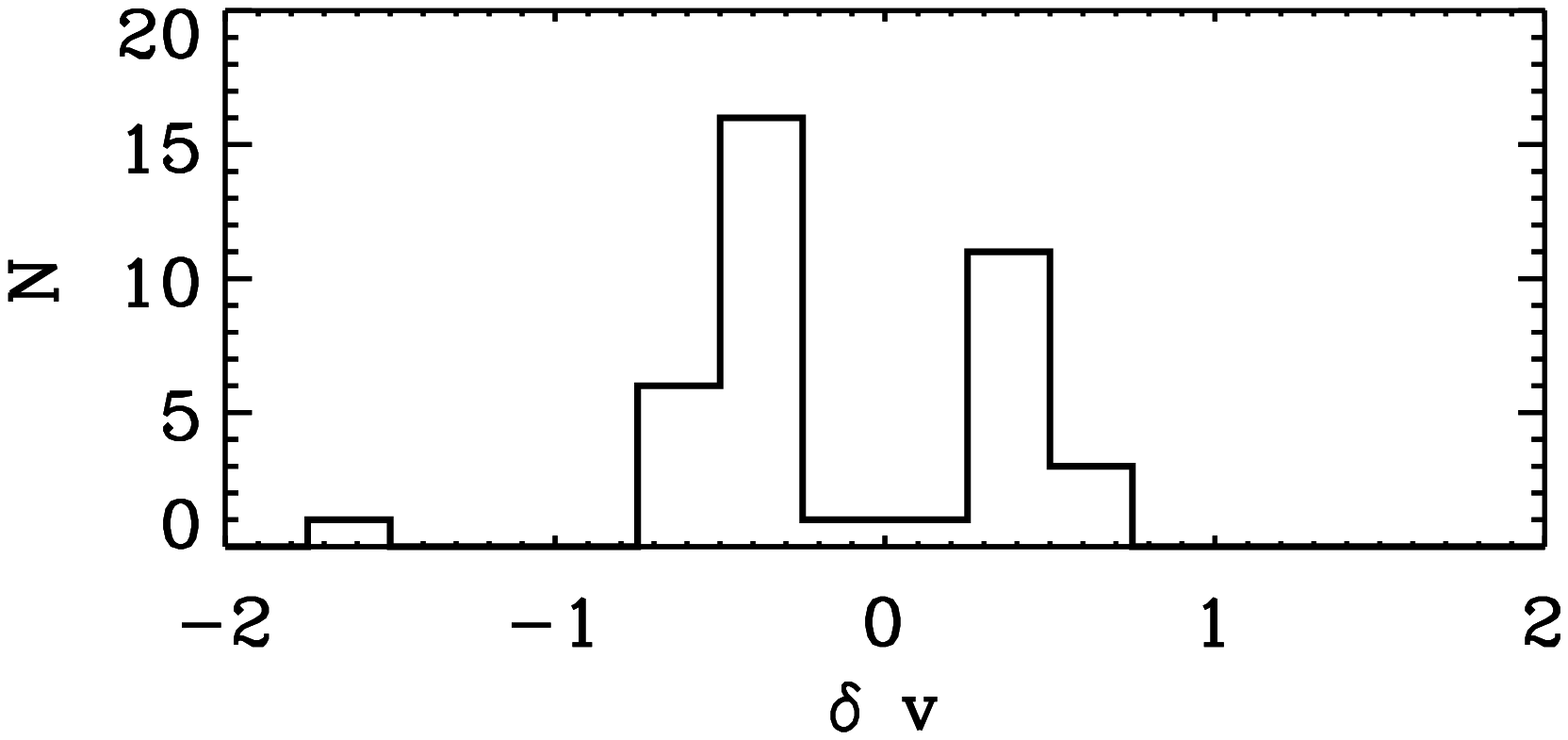}\\
\includegraphics[width=3in]{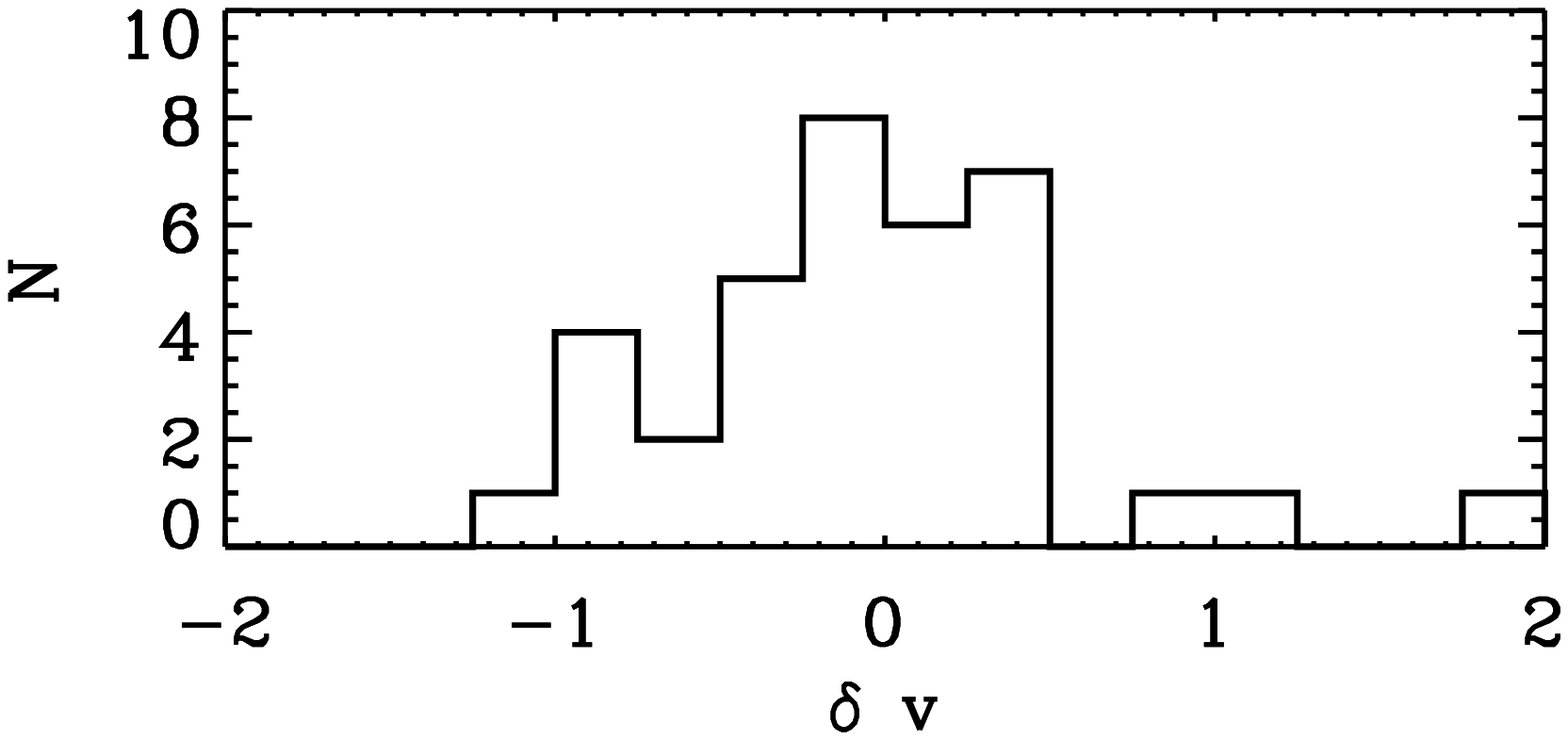}
\caption{\textit{Top} The normalised velocity difference between the HCN F(2-1) line and isolated \n(1-0) line. \textit{Bottom} As above but for the CS(2-1) line. Both distributions, but particularly the HCN, are skewed to the blue side of the distribution indicative of infall motions.}
\label{dV}
\end{center}
\end{figure}

An alternative method of classifying cores is to use the normalised velocity difference $\delta V$ between the optically thick and thin lines \citep{Mardones97}. This is calculated by the equation
\begin{equation}
\delta V = (V_{thick}-V_{thin})/\Delta V_{thin}
\end{equation}
where $\Delta V_{thin}$ is the full width half maximum (FWHM) of the optically thin line. We calculate this value for the isolated HCN hyperfine line and for the CS(2-1) lines by fitting a Gaussian profile to the \n\ line to obtain the line centre and FWHM. We then find on which side of the \n\ line centre the optically thick tracer peaks and fit a Gaussian to this component to obtain the optically thick component. \fig \ref{dV} shows the results of these calculations.

A negative value of $\delta V$ is blue shifted and a positive value red shifted. An observation is considered to be a detection of true infall or expansion if $| \delta V |$ is greater than $5 \sigma(\delta V)$. For HCN, 20/42 (48\%) of the line profiles are classified as blue shifted collapse profiles, and 13/42 (31\%) of the line profiles are red expansion profiles. This is slightly higher than in the above classification by profile shape as some of the profiles previously classified as ambiguous are now classified as collapsing or expanding. It is interesting to note in \fig \ref{dV} that there is a gap in the distribution around the zero point. This could be either due to statistical effects or that our sample is biased as it only contains dynamical collapsing cores and not any static cores.

The $\delta V$ distribution for CS(2-1) is very slightly skewed towards the blue side. For CS there are 16/42 (38\%) profiles where $\delta V$ is indicative of collapse and 14/42 (33\%) where $\delta V$ is indicative of expansion. In addition to the high false positive rate of red profiles, many of the Gaussian fits to the CS(2-1) lines are too poor for the core to be classified. As in the case of the above classification by profile shape, the CS(2-1) line is a poor indicator of collapse for cores embedded in dense dynamical filaments. Given these detection statistics, it is clear that the velocities that the CS traces are largely unrelated to those within the dense core. Subsequently, we shall focus on the HCN lines. As such we are taking the optimal case for detecting a blue asymmetric line profile from a collapsing core.

All of the filaments contain a collapsing core, yet in more than 50\% of cases, in even the most sensitive tracer (HCN), this collapse does not reveal itself as a blue asymmetry in the line profile. This leads us to the first conclusion of this study: filaments can obscure signatures of collapse from their embedded cores.

\subsection{Velocity structure of the filament}\label{velocities}

\begin{figure*}
\begin{center}
\begin{tabular}{c c}
\includegraphics[width=3.5in]{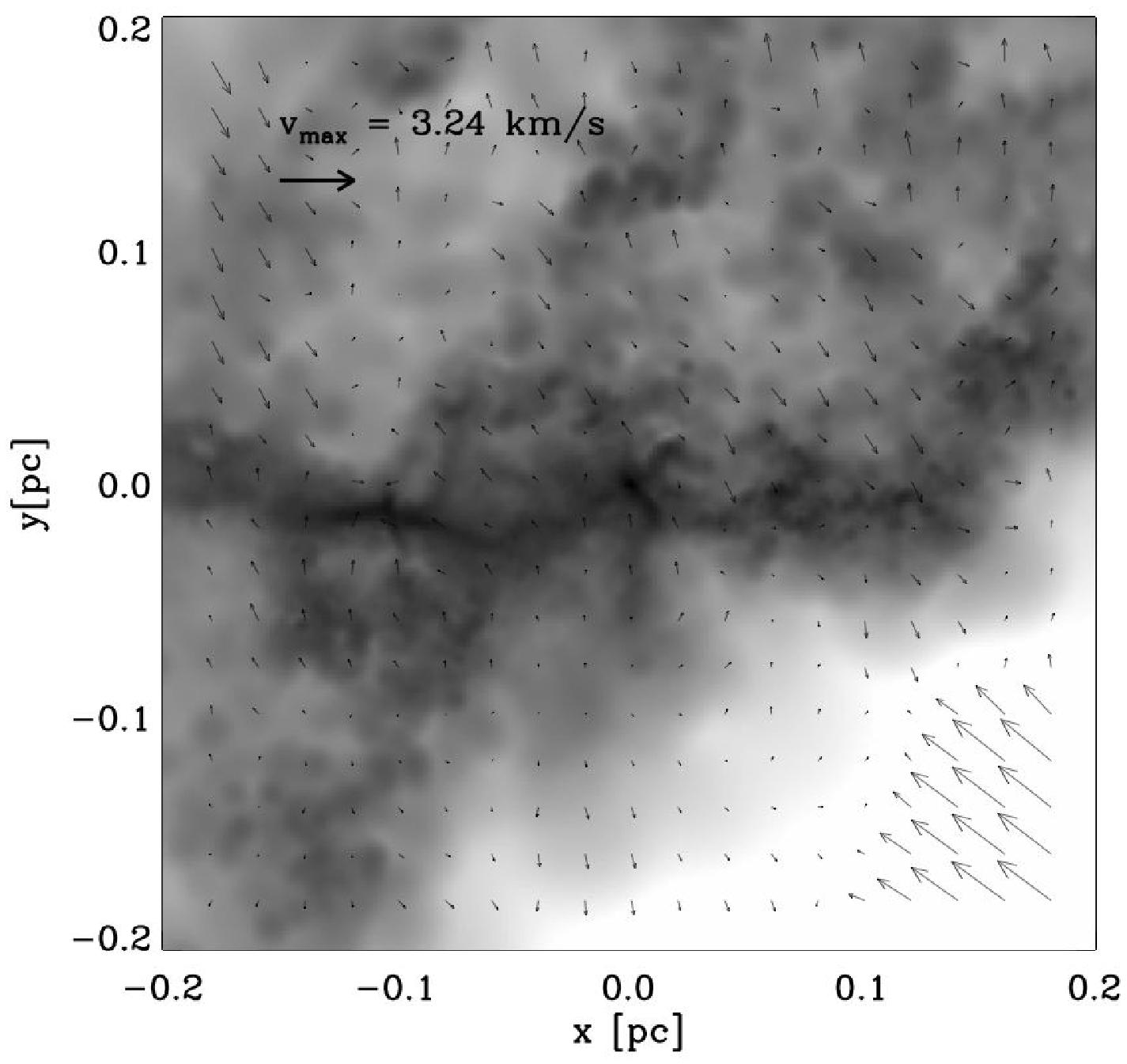}
\includegraphics[width=3.5in]{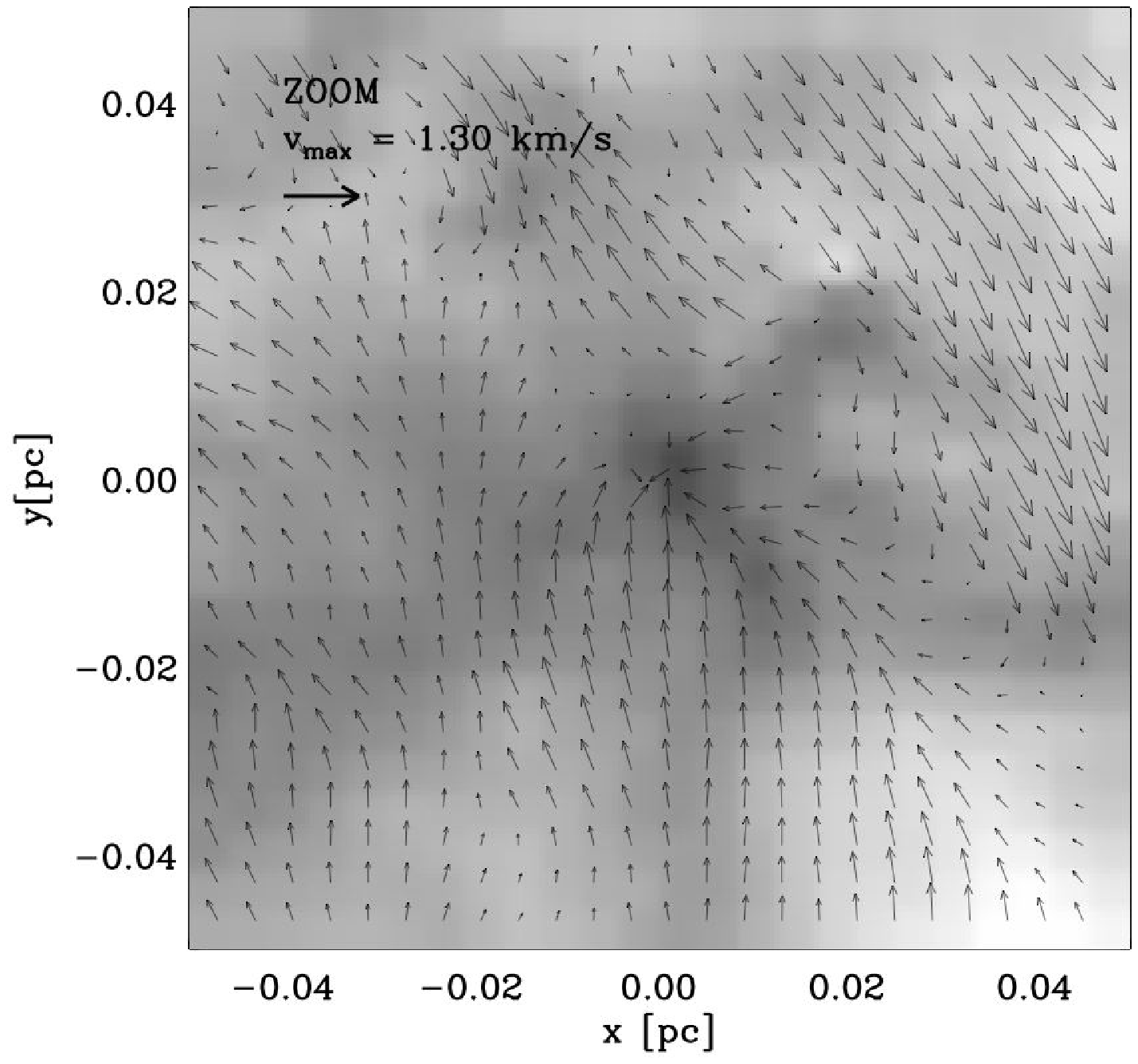}\\
\includegraphics[width=3.5in]{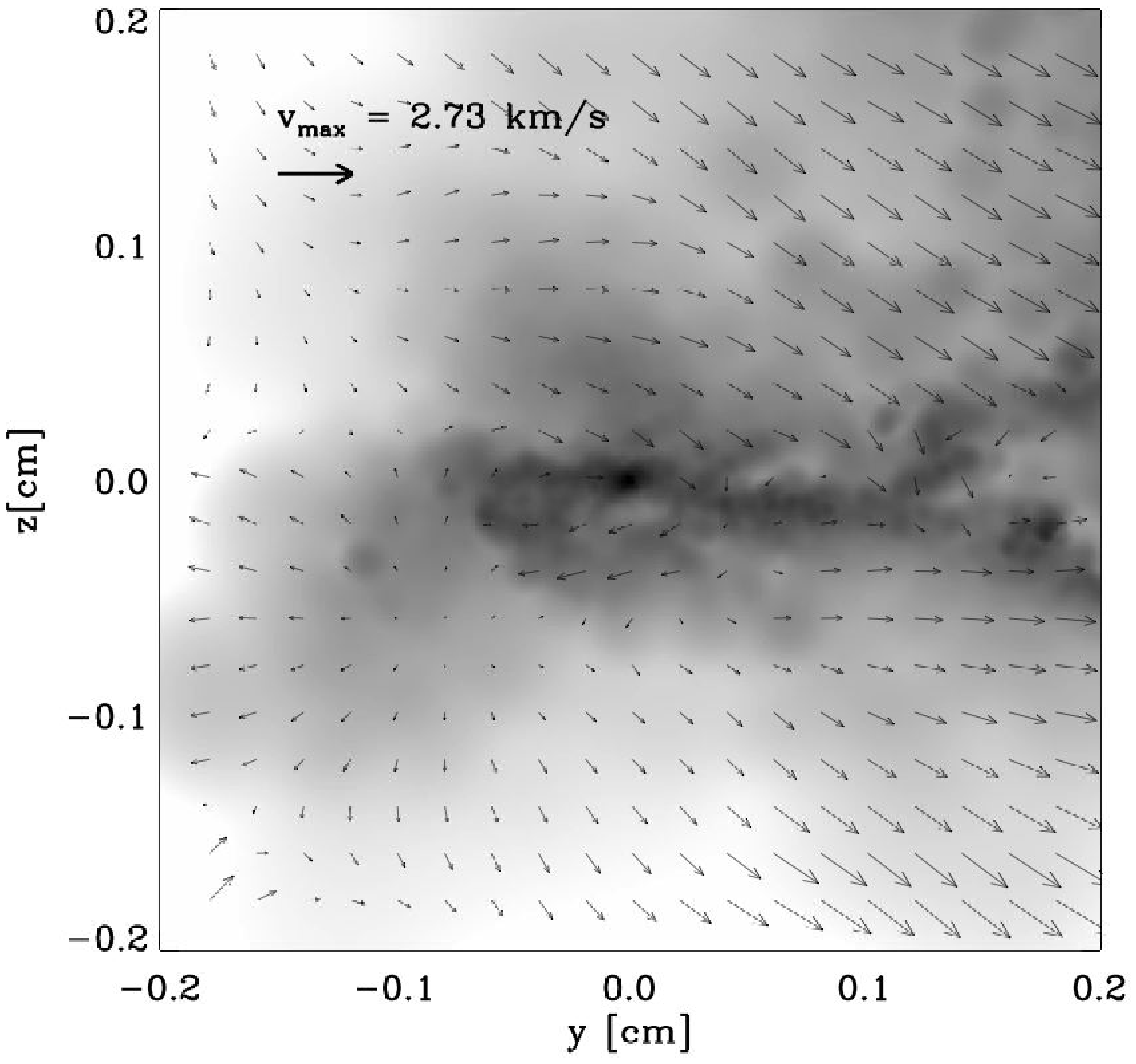}
\includegraphics[width=3.5in]{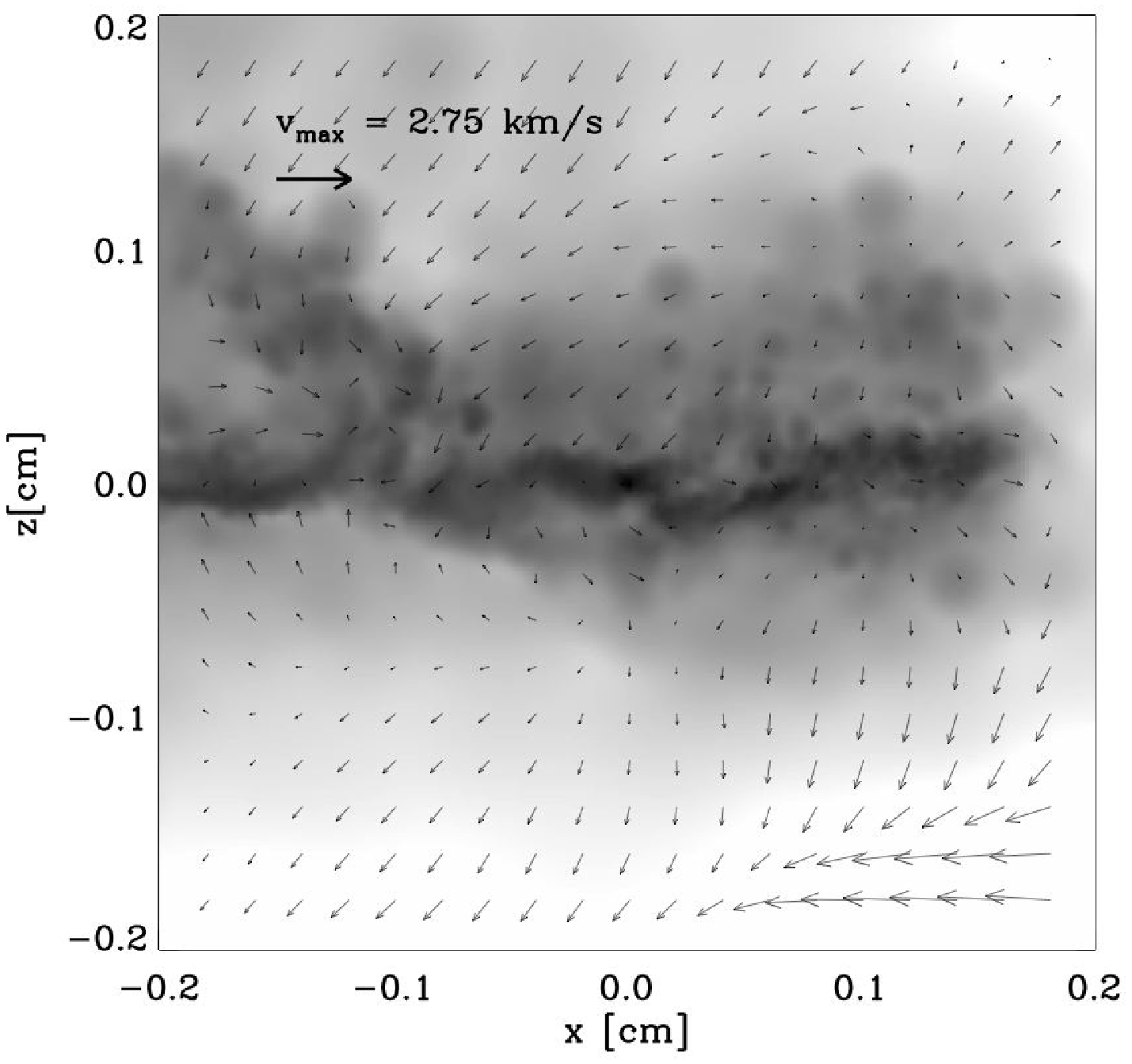}\\
\end{tabular}
\caption{The velocity vectors of material in three perpendicular slices through the centre of the embedded core in our original model of Filament A before radiative transfer. The background greyscale shows the densities in the slice (which has a depth of 0.002 pc.)}
\label{velvectors}
\end{center}
\end{figure*}

Given the importance of the dynamical state of the envelope on the resulting core line profile, we examine in more detail the dynamics of the filaments. \fig \ref{velvectors} shows the velocity field in three slices that intersect the core in Filament A, and a zoom into the centre of the x-y plane. Filament A is a turbulent sheet with a disordered velocity field within the x-y plane. There is no systematic velocity along the filament's major axis, and there are several regions that are contracting. In the y-z and x-z planes the diffuse material surrounding the filament has a clearer systematic velocity gradient and a larger absolute velocity. 

Sheets and filaments are well known to be one of the main sites where turbulent energy is transferred from large scales to small. For example, \citet{Boldyrev02} extended the \citet{She94} model of incompressible turbulence to the supersonic compressible regime and found that making the assumption that dissipation occurred in micro-turbulent sheets achieved a good agreement with numerical simulations. More recently, \citet{Schmidt08} analysed high resolution turbulent boxes with various forcing scales and found that the most dissipative structures are intermediate between filamentary and sheet-like structures. Therefore it is to be expected that shocks will dissipate energy and transform turbulence from high amplitude large scale motions in the diffuse gas into many low amplitude small scale motions within the denser filament. The velocities shown in \fig \ref{velvectors} are therefore a natural consequence of filament formation from a turbulent molecular cloud.

The zoom in to the central core of Filament A shown in \fig \ref{velvectors}c shows the detailed velocity structure surrounding the central embedded core. In this case the flow onto the central core is not radially symmetric, but instead there are twisting streams of material flowing onto the core. Behind the core there is even a stagnation point in the velocity field (similar effects have been seen in \citealt{Ballesteros-Paredes03} and \citealt{Klessen05}). Any sight line which passes through this stagnation point will differ substantially from the theoretical velocity profile shown in \fig \ref{schematic}.

\begin{figure*}
\begin{center}
\begin{tabular}{c c}
\includegraphics[width=3.5in]{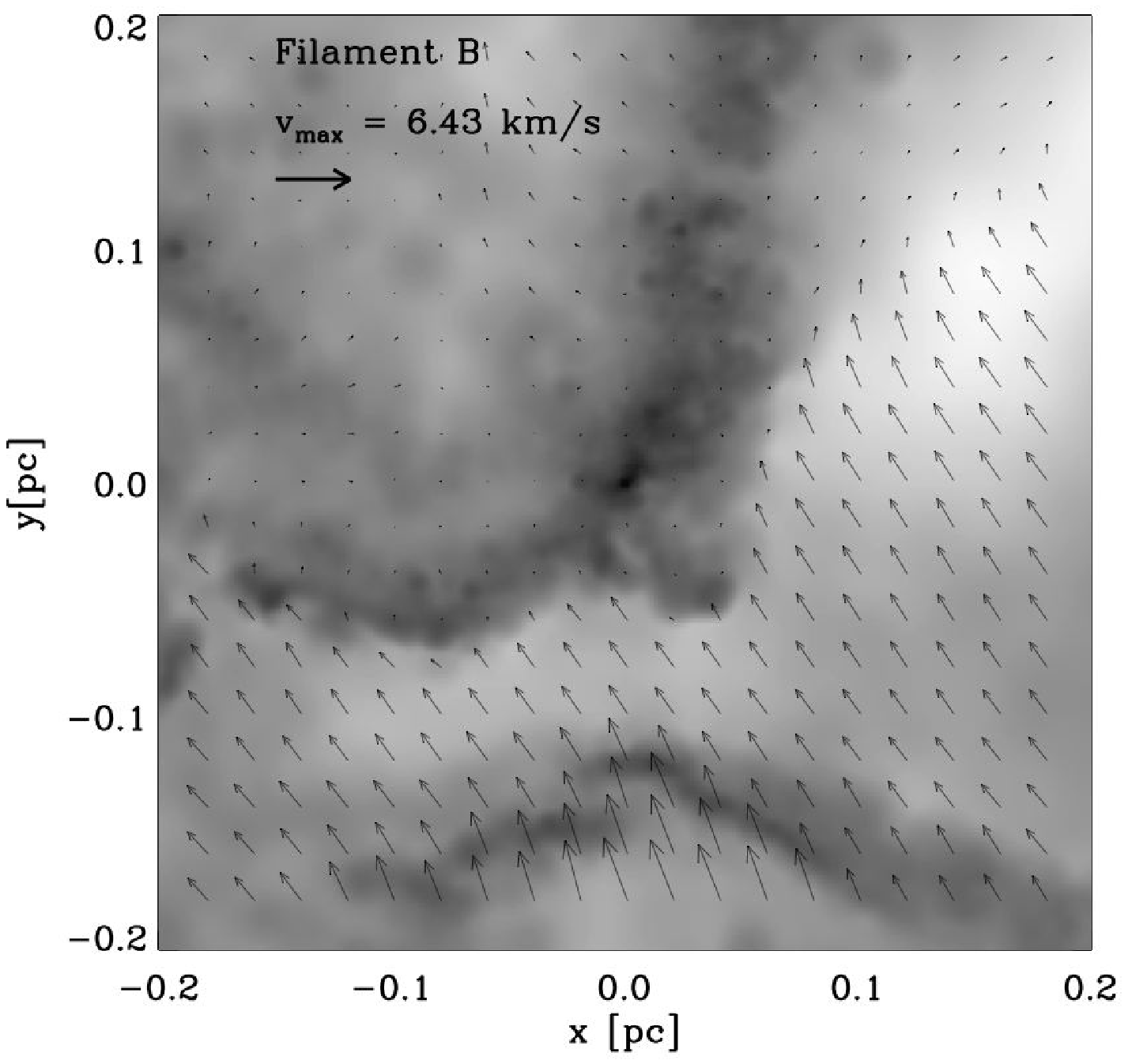}
\includegraphics[width=3.5in]{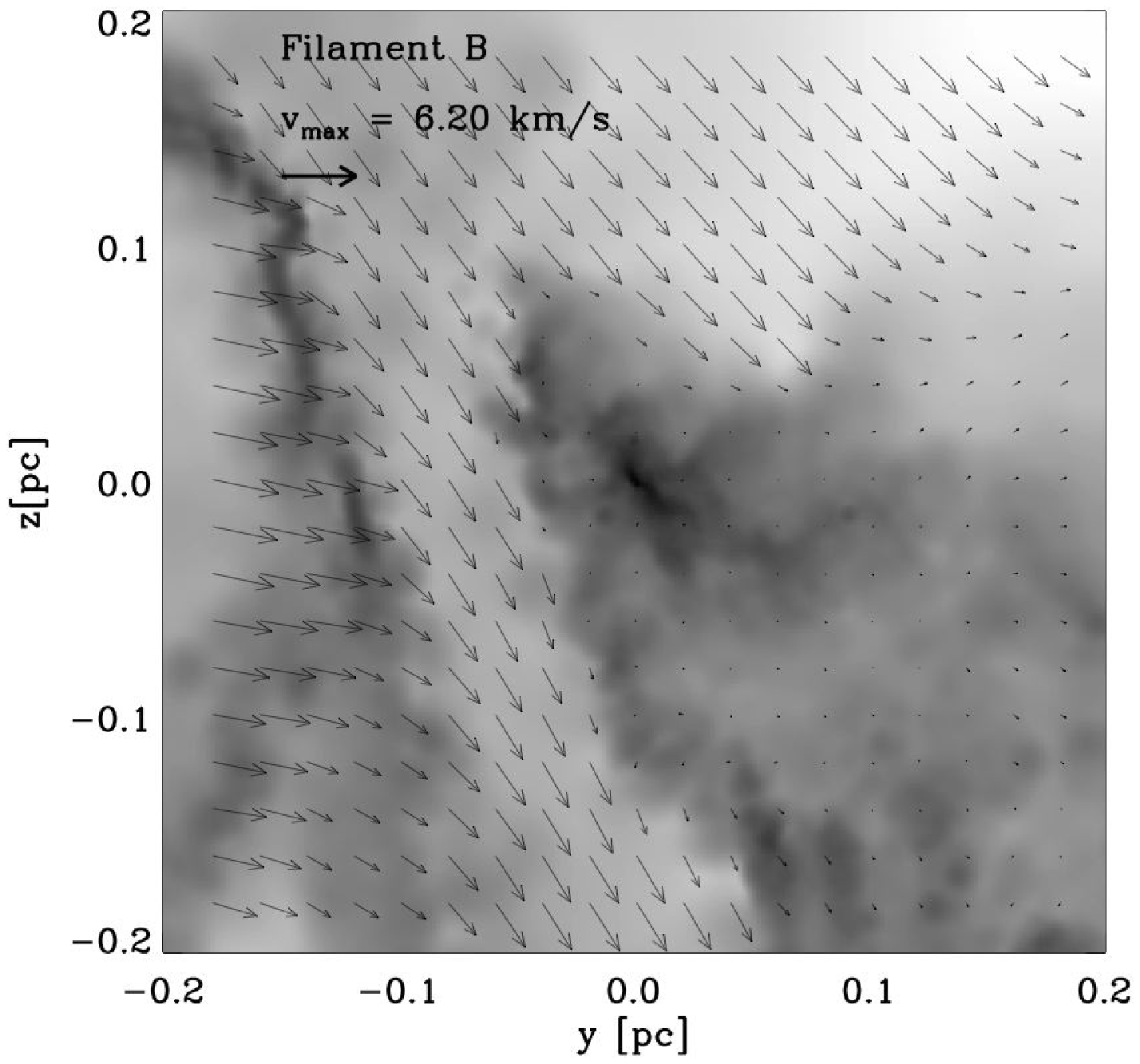}\\
\includegraphics[width=3.5in]{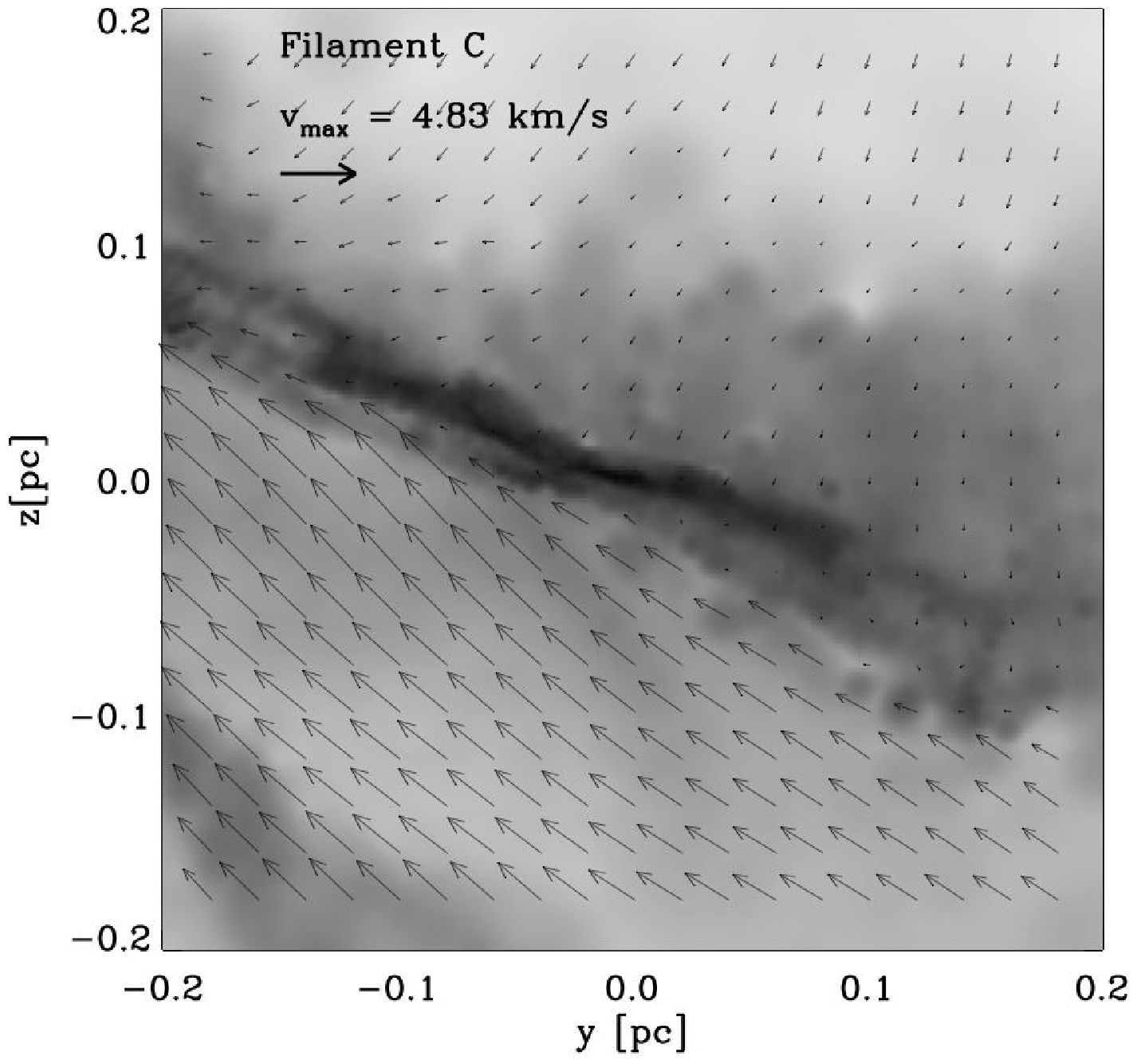}
\includegraphics[width=3.5in]{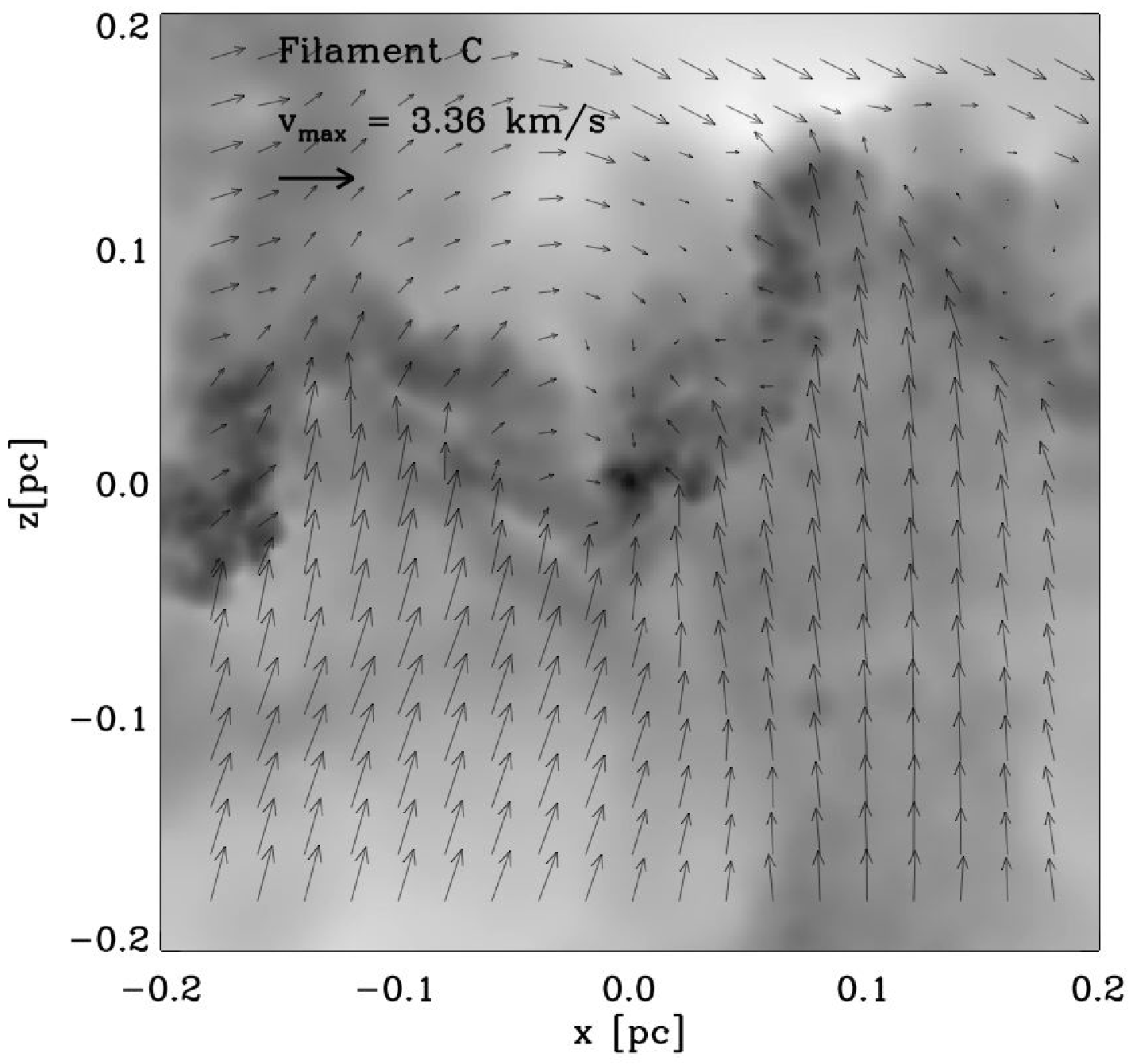}\\
\end{tabular}
\caption{The velocity vectors in slices through the central core in our model of Filament B \textit{top} and Filament C \textit{bottom}. The background greyscale shows the densities in the slice. The filaments are formed at the interface of converging flows.}
\label{velvectors2}
\end{center}
\end{figure*}

\fig \ref{velvectors2} shows a slice through Filaments B and C in a similar manner to \fig \ref{velvectors}. In Filament B a fast moving flow of gas is pushing against a denser region of gas with lower mean velocities. The dense filament is found at the interface of the two regions and along this boundary turbulent whorls of gas can be seen. Along this interface material from the fast-moving diffuse flow will shock onto the surface of the filament. Filament C is once again formed at the interface between two flows. In this case, however, the contrast is even more extreme as the flows are converging. Both flows are supersonic which will lead to a rapid accumulation of dense material at their interface.

Clearly the velocities in the filaments within which the cores are embedded are very different from the quiescent enveloped envisage in \sect \ref{theory}. Moreover, these velocity flows are an intrinsic part of the filament as they result as a byproduct of its formation. Indeed it has been proposed that gravitational contraction and accretion are drivers of turbulence on virtually all scales \citep{Klessen10}. Many previous discussions of filamentary geometries in the literature have focussed on either hydrostatic filaments \citep{Ostriker64} or their subsequent self-similar collapse \citep{Hennebelle03,Tilley03}. The filaments in our large GMC simulations are dynamic objects that form as a result of turbulence and the associated shocks, and therefore have a much more disordered velocity field (see also \citealt{Klessen00b}).

\begin{figure*}
\begin{center}
\begin{tabular}{c c c}
\includegraphics[width=2.35in]{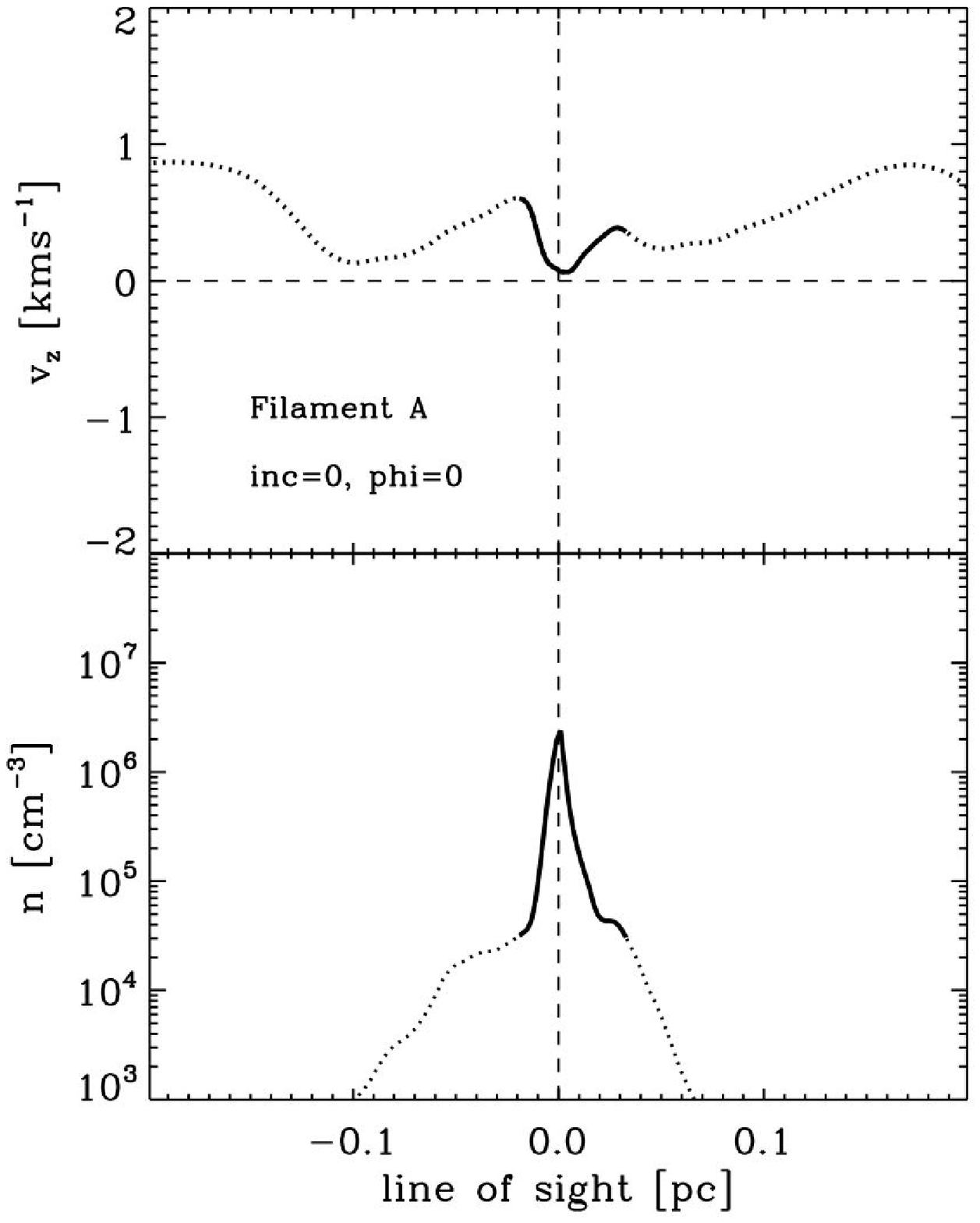}
\includegraphics[width=2.35in]{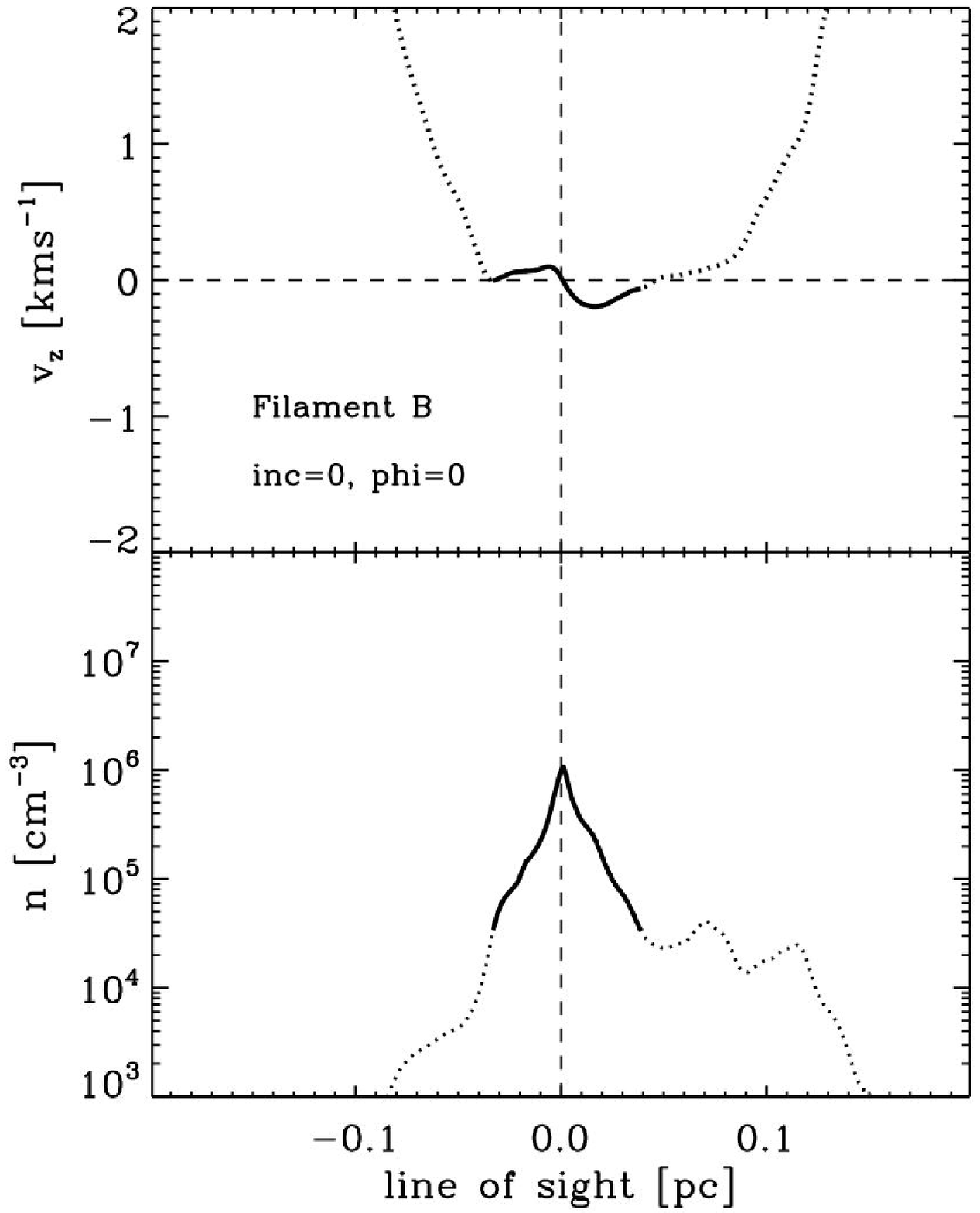}
\includegraphics[width=2.29in]{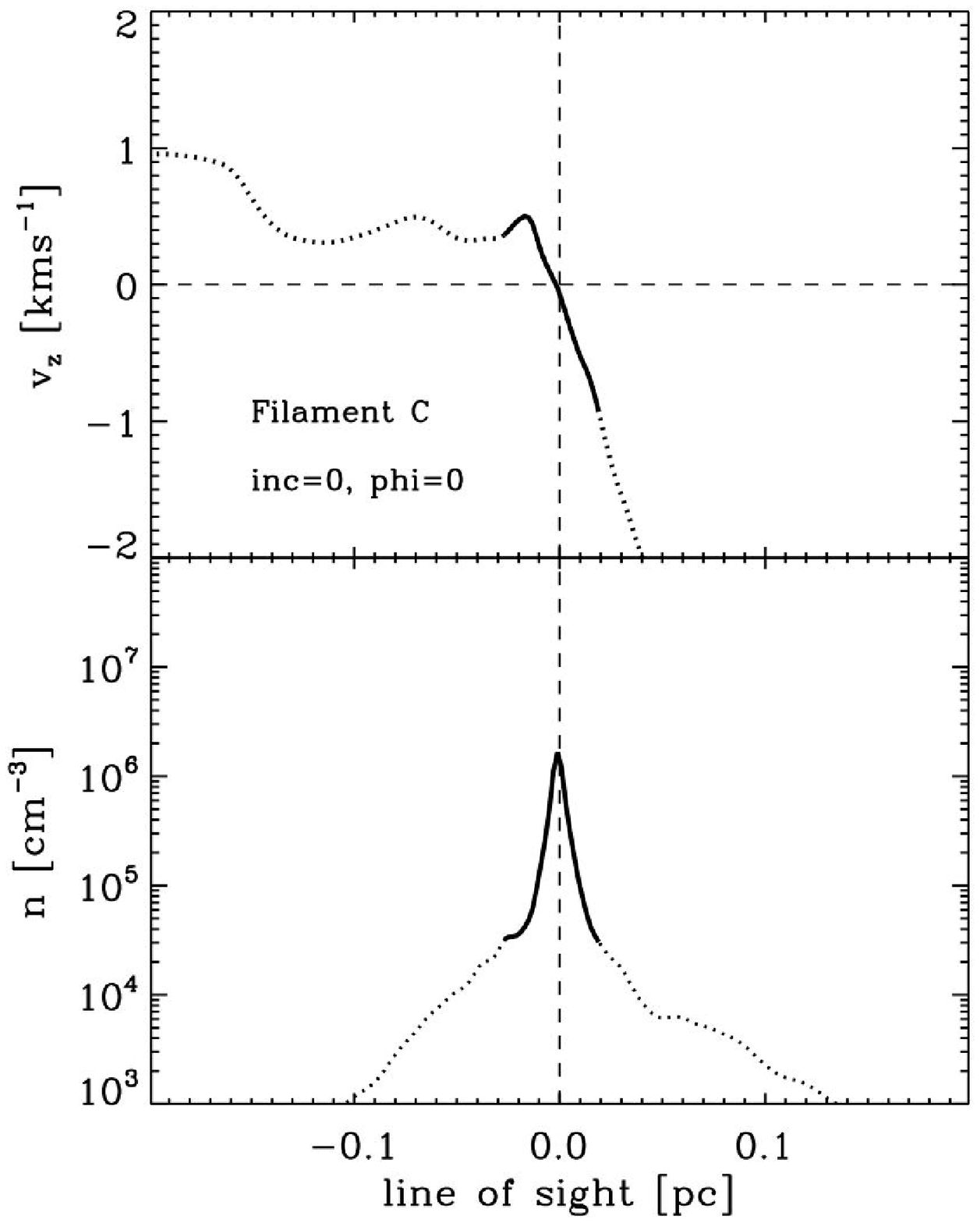}
\end{tabular}
\caption{The line of sight properties averaged over our Gaussian beam along the inc=0\degree\ phi=0\degree\ direction. The line is solid in the core region (taken as a density above $3\E^4$ \cmc) and dotted where the gas is more diffuse. The velocity and density components along the line of sight are more complex than the simple symmetrical model shown in \fig \ref{schematic}.}
\label{los_properties}
\end{center}
\end{figure*}

The effect that these disordered velocities in the plane of the filament have on the line profiles is illustrated in \fig \ref{los_properties}, indicating the velocities and densities of the material along the inc=0\degree,  phi=0\degree\ line of sight (this corresponds to the -z axis shown in Figures \ref{velvectors} and \ref{velvectors2}). This figure provides examples of how the physical conditions in Figures \ref{velvectors} and \ref{velvectors2} gave rise to the line profiles shown in \fig \ref{sightlines}.

The left panel of \fig \ref{los_properties} shows Filament A. Along this line of sight the stagnation point is behind the core, so there are no blue-shifted velocities. In other words, most of the mass the core accretion occurs from one direction. Consequently, when viewed from the forwards direction (inc=0\degree), the densest visible peak in the velocity profile is on the blue side and when viewed from the the reverse direction (inc=180\degree) the densest visible peak is on the red side. This gives rise to a profile which is either red or blue asymmetric depending on the viewing angle.

The central panel of \fig \ref{los_properties} shows Filament B. Here the velocity profile in the core is similar to that observed in \fig \ref{schematic}. However, the region as a whole exhibits a large scale velocity gradient as the filament is swept along with a larger external gas flow. On the forward (inc=0\degree ) side, the optically thick emission comes from the densest point of the blue-shifted velocities, and a blue asymmetric profile is produced. However on the reverse side there is optically thick emission at the same velocity in front of the core. This obscures the blue-shifted core emission, and so no blue asymmetry is seen.

The right panel of \fig \ref{los_properties} shows Filament C. In this filament the velocity profile of the collapsing core joins smoothly on to the large scale convergent motions that are forming the filament. Here the converging gas flows first bring the filament together, and then at higher densities gravity leads to the formation of a dense core (see also \citealt{Hennebelle08,Banerjee09}). When viewed from the reverse direction (inc=180\degree) the gas distribution in front of the core does not overlap in velocity space with the central core. Consequently all the optically thick emission from the core is visible to the observer, and there is neither a self absorption dip at the centre of the line profile, nor any asymmetry between the blue and red side of the line.

\subsection{Line-widths}
\begin{figure*}
\begin{center}
\begin{tabular}{c c c}
\includegraphics[width=2.1in]{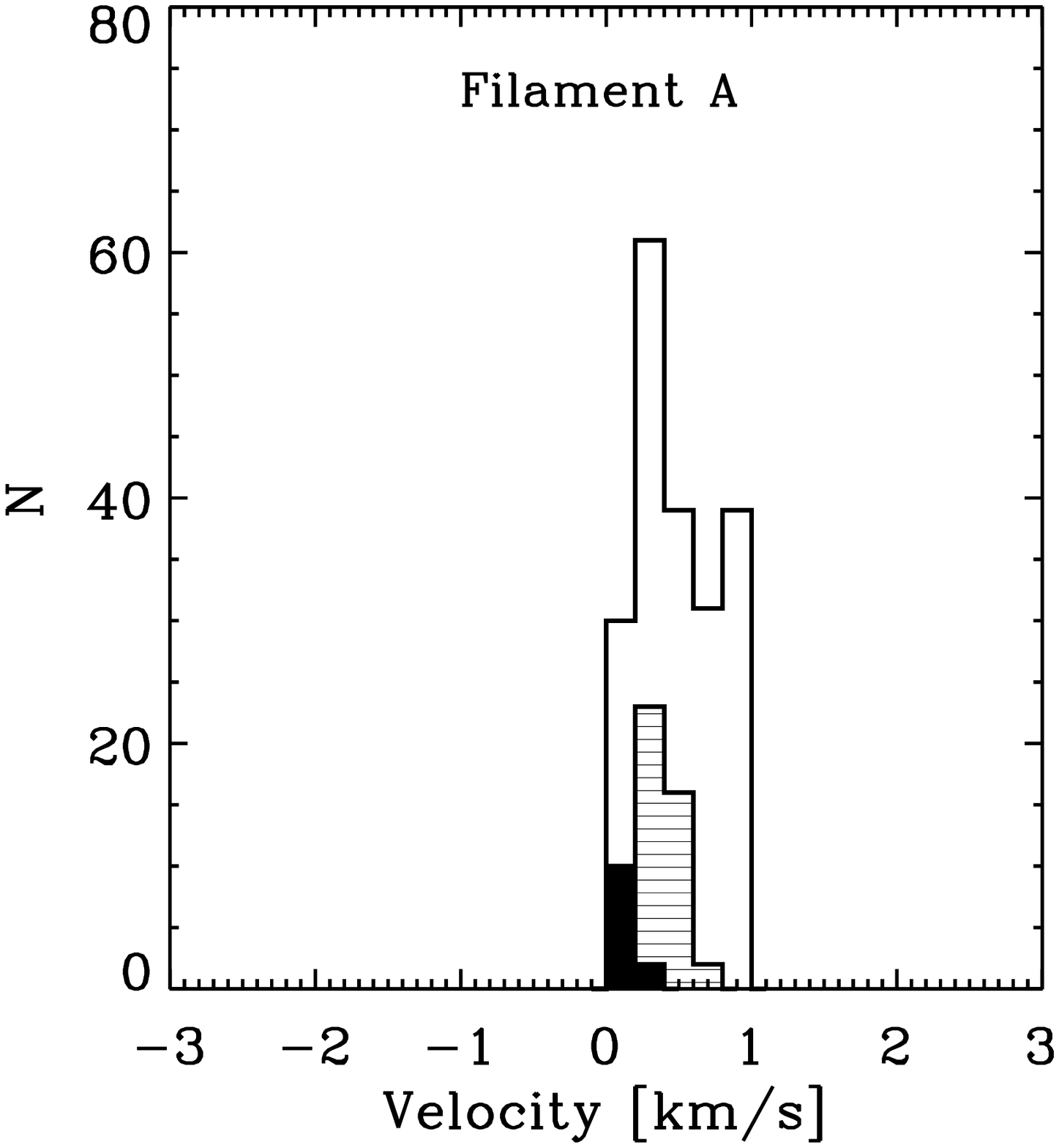}
\includegraphics[width=2.1in]{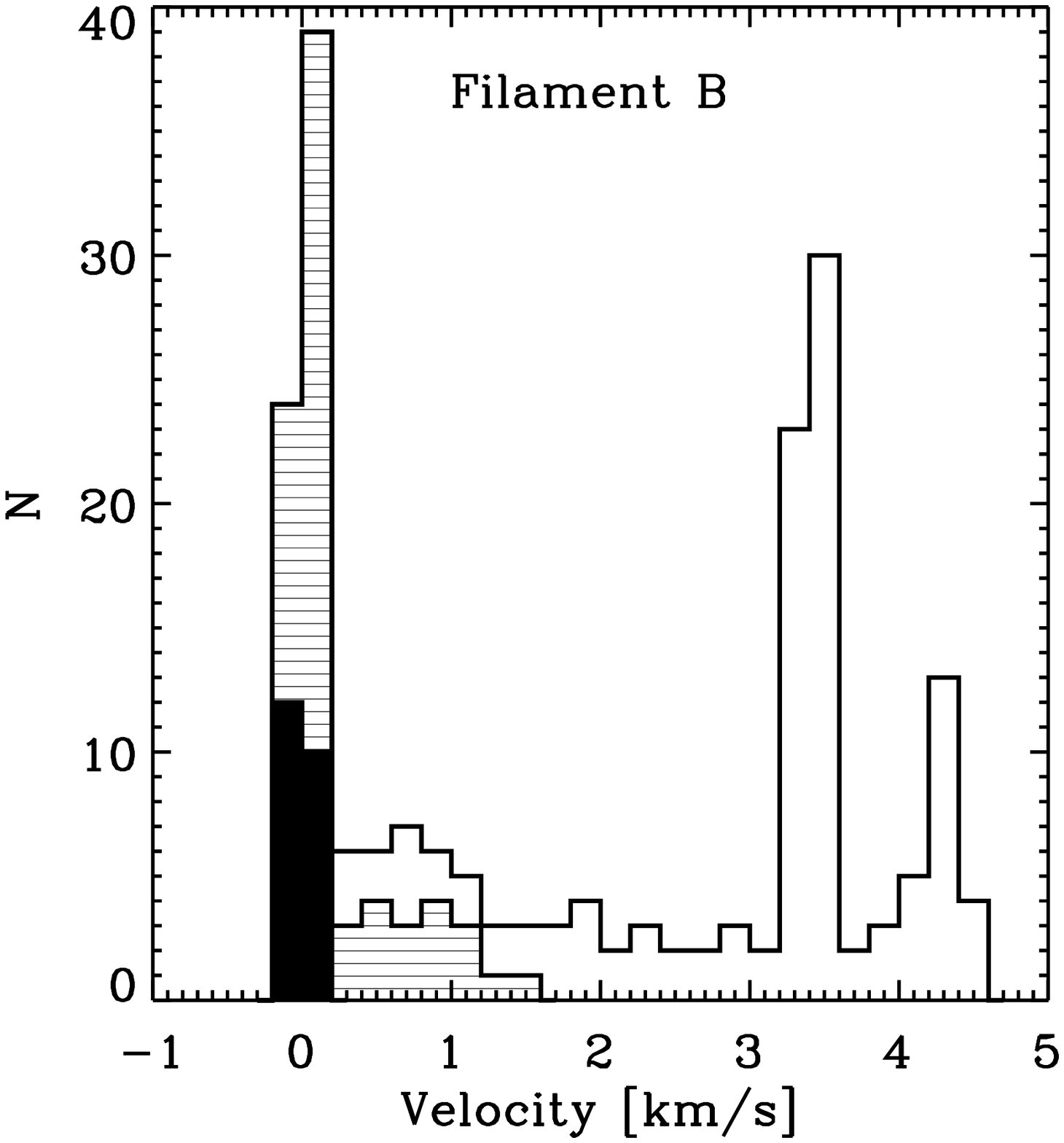}
\includegraphics[width=2.1in]{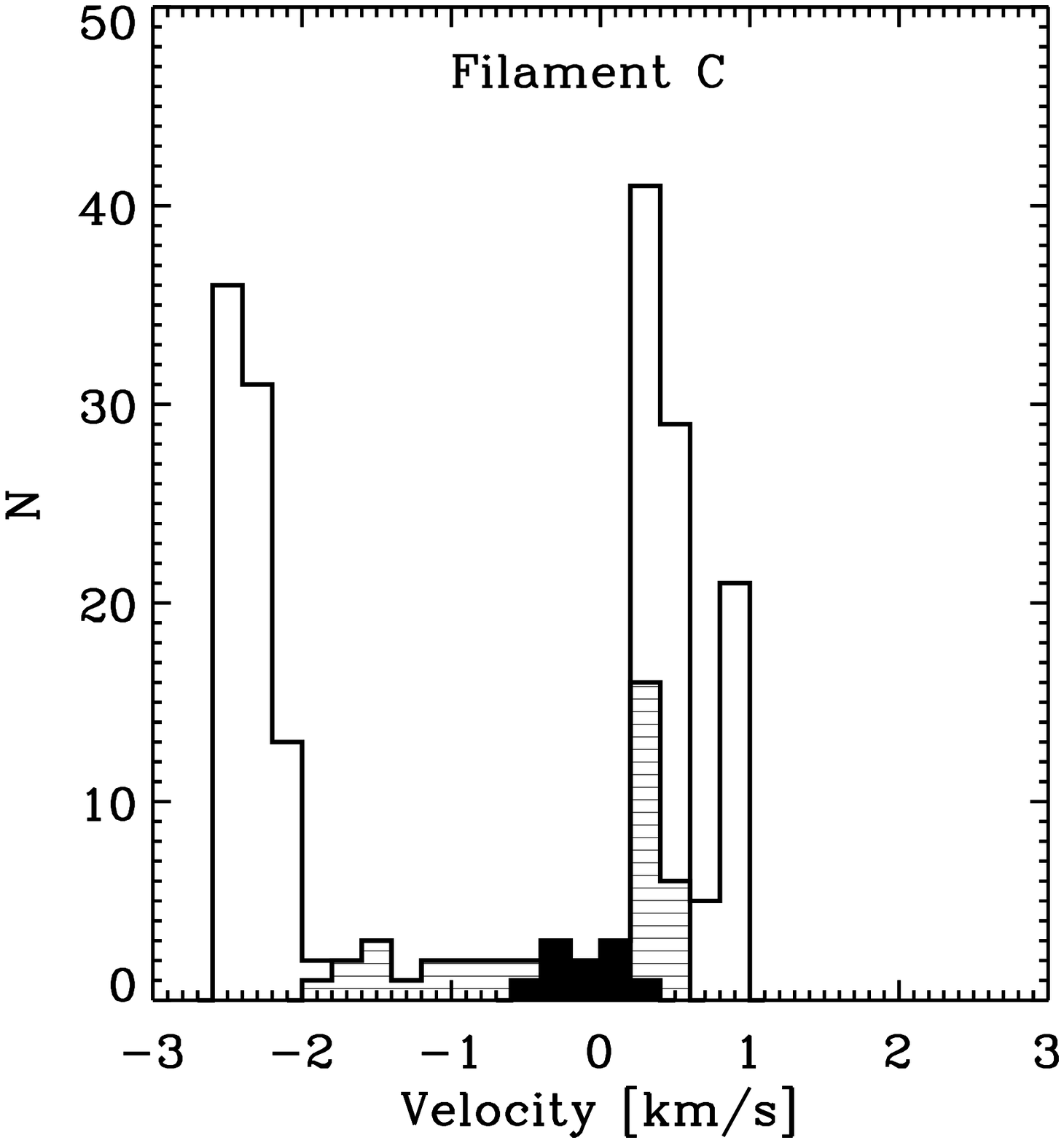}\\
\end{tabular}
\caption{Histograms of the distribution of velocities which make up the Gaussian beam along the inc=0\degree\ phi=0\degree\ direction for the filaments. The velocities in the clear histogram are sampled at equally spaced intervals along the beam. The patterned histogram contains only those points within the beam with a density above $10^4$ \cmc, and the solid histogram contains only those points with a density above $10^5$ \cmc. The underlying velocity distributions are non-Gaussian and reflect the filaments formation from bulk gas flows. However, the densest material typically occupies a narrow, typically sonic, region within this velocity space.}
\label{vhist}
\end{center}
\end{figure*}

Another interesting diagnostic is the linewidths of the \n\ profiles. \tab \ref{linewidths} shows the mean velocity dispersion, $\sigma(v)$, of the \n\ lines when fitted by the following Gaussian function
\begin{equation}
G=A_0 \exp\left[\frac{-(v-v_0)^2}{2\sigma^2(v)}\right]
\end{equation}
where $A_0$ is the peak line brightness and $v_0$ the rest velocity. \tab \ref{linewidths} also lists the velocity dispersion along the beam directly from the model. We calculate both the volume weighted and density weighted velocity dispersions. The volume weighted dispersion is larger than the density weighted value, as expected from the fact that the core forms at the stagnation point of a convergent flow. To illustrate this, \fig \ref{vhist} shows a histogram of the simulation velocities at equally spaced intervals along the beam for the filaments. In the case of Filaments B and C the velocity distribution is non-Gaussian and appears bimodal, which reflects the converging flows of gas around the filament. In \fig \ref{vhist} we also plot two subsets of the data; points with densities above $10^4$ \cmc, and points above $10^5$ \cmc. Both subsets occupy a narrower velocity range than the parent distribution, and the velocity range of the $10^5$ \cmc gas which is associated with the cores typically is around the sonic scale. This explains why the `observed'  \n\ linewidths of the cores are roughly sonic.

\begin{table}
	\centering
	\caption{The mean line-width of the \n\ line compared to the velocity dispersions in the simulation. The velocity dispersion of the line $\sigma(v)$ is found from a Gaussian fit to the line profile from each viewing angle, which is then averaged to find the mean. The velocity dispersion of the simulation is found by calculating the beam weighted mean velocity and its standard deviation along each sightline, and then calculating the average over all the sightlines. As well as the volume weighted average we also calculate the density weighted average. The linewidths are roughly sonic despite the volume weighted velocity dispersions being supersonic.} 
		\begin{tabular}{l c c c}
   	         \hline
	         \hline
	         Filament & Mean $\sigma(v)$\\
	         & Line Profile & Volume Weighted & Density weighted\\
	         \hline
	         A  & $0.28\pm0.07$ & $0.45\pm0.31$ & $0.10\pm0.08$\\
	         B  & $0.20\pm0.02$ & $0.94\pm0.67$ & $0.39\pm0.36$ \\
	         C  & $0.20\pm0.05$ & $0.91\pm0.47$ & $0.50\pm0.24$ \\
		\hline
		\end{tabular}
	\label{linewidths}
\end{table}

That the velocity dispersions found from the \n\ lines are typically coherent is in good agreement with recent observational studies of star-forming filaments. \citet{Pineda11} carried out NH$_3$ observations of a filament in the Barnard 5 region of Perseus and found it to be internally coherent, with a sharp transition to supersonic motions at its boundary. \citet{Hacar11} observed filaments with embedded cores in L1517. They found that both the embedded cores and the dense gas in the filaments had subsonic velocities when observed in C$^{18}$O, SO and \n. The filaments in our models are formed from supersonic shocks. Before the shock (on the outside of the filament) the velocities are supersonic and the gas is diffuse, after the shock, the velocities are subsonic and the gas is dense in good agreement with these observations. In general dense cores are coherent objects \citep[e.g.][]{Goodman98,Kirk07}. An important feature of this simulation is that we produce coherent cores purely by hydrodynamics without having to consider magnetic fields (see also \citealt{Ballesteros-Paredes03,Klessen05}).

The linewidths of the optically thick species appear broader than the \n\ linewidths. Both \n\ and HCN have similar Einstein coefficients and high critical densities. Consequently both species are tracing similar gas; therefore this effect cannot be attributed to the underlying velocity field, but must instead be caused by optical depth effects. As the HCN emission is optically thick, emission from the dense core is obscured by the surrounding gas, which reduces the intensity of the central peak. The line wings, however, are less obscured and therefore make up a greater part of the total emission than in the optically thin case, explaining why the optically thick lines appear flatter and broader. In the case of CS freeze-out this effect is compounded in the dense core, which further increases the relative contribution of emission from higher velocity gas in the filamentary envelope.

\subsection{The surrounding line profiles}
\begin{figure}
\begin{center}
\includegraphics[width=3.2in]{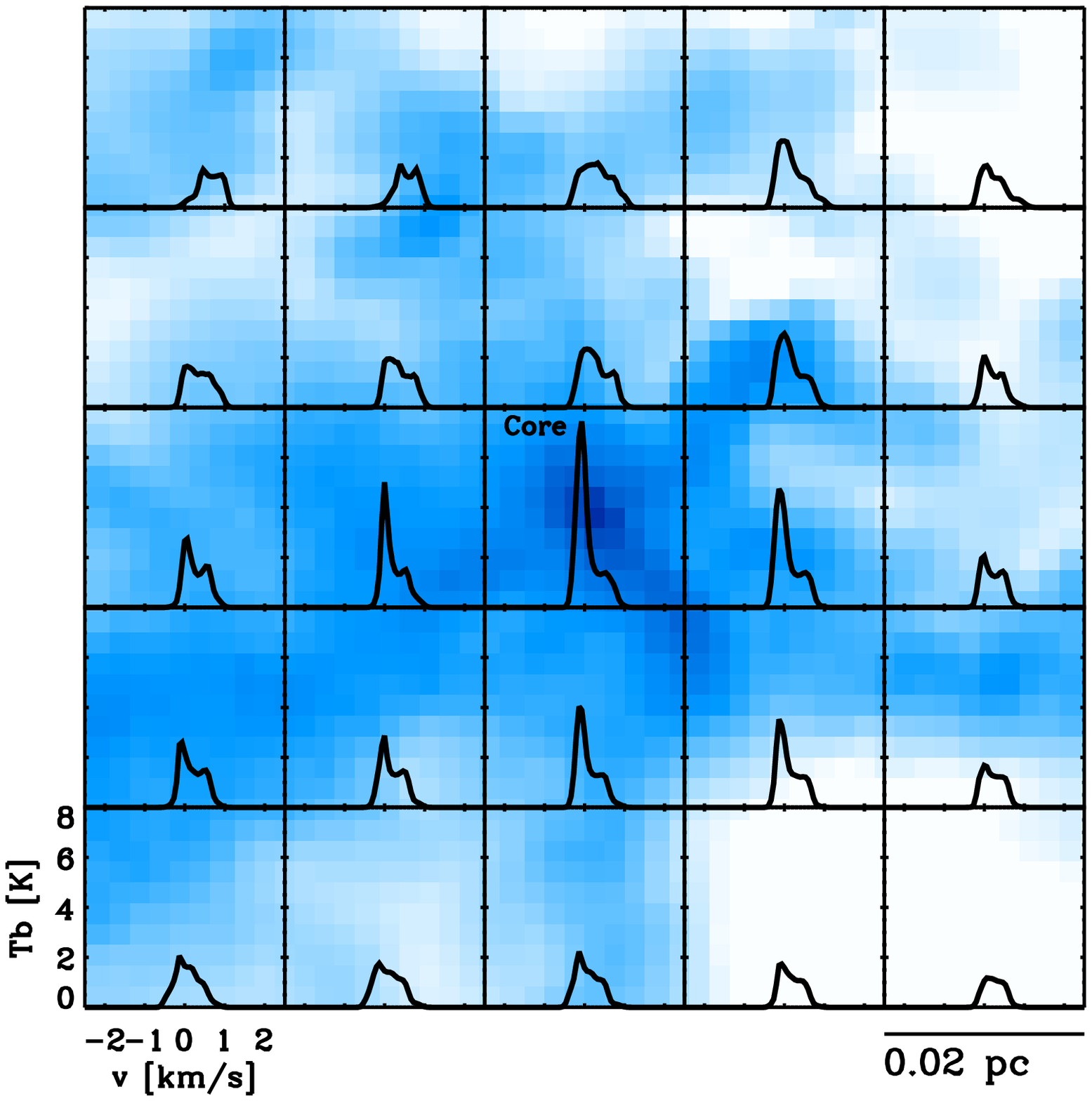}\\
\includegraphics[width=3.2in]{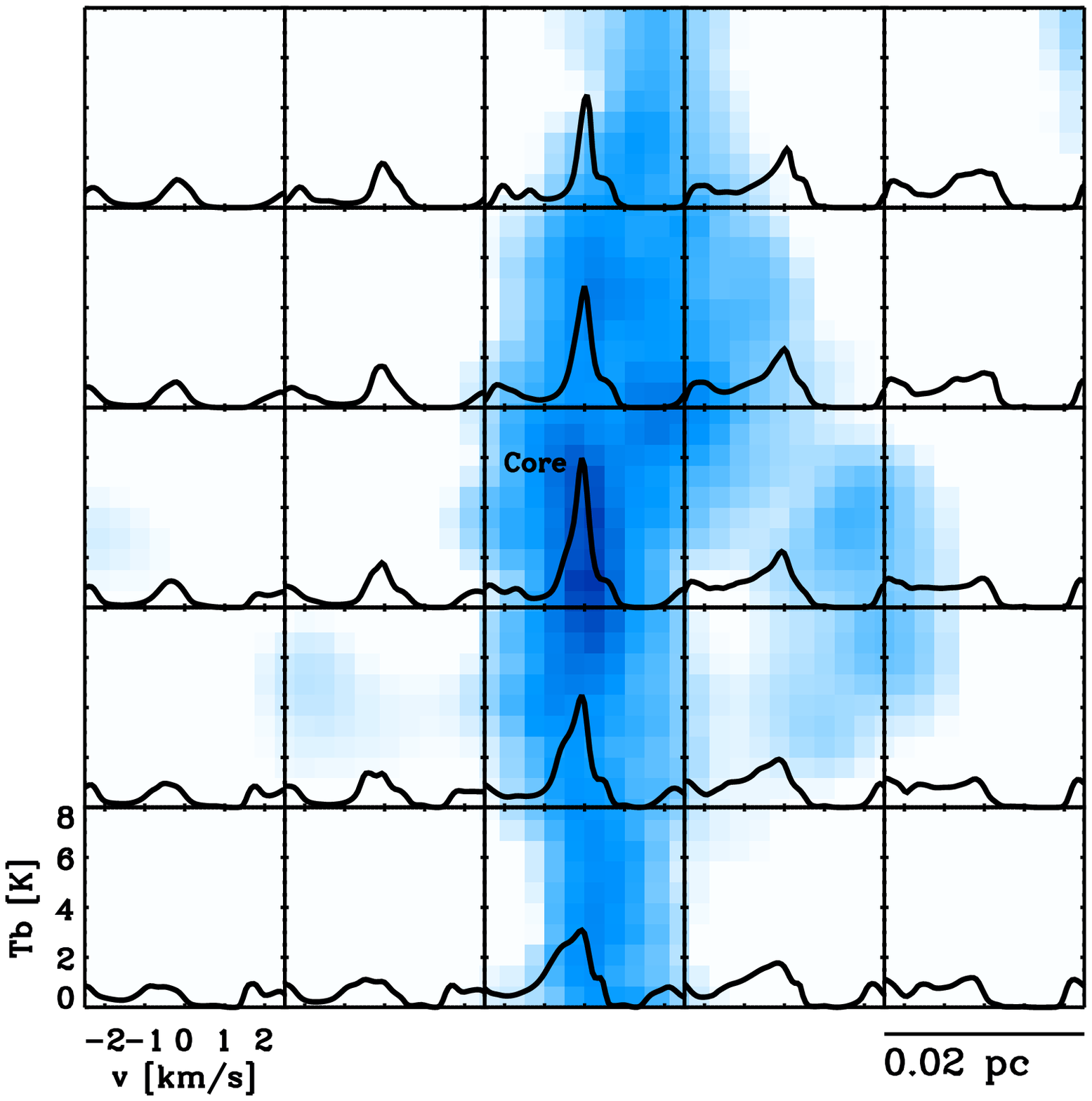}\\
\caption{The HCN F(2-1) lines of the region surrounding the central core of Filament A \textit{(left)} and C \textit{(right)}. The background image shows the dust emission from the filaments.}
\label{surround}
\end{center}
\end{figure}

For a core to be interpreted as collapsing, its line profile should differ from its surroundings. \fig \ref{surround} shows the HCN F(2-1) line profiles of the regions surrounding the central cores embedded within Filaments A and C. The cores are shown at an inclination and rotation of zero where each filament has a blue asymmetric line profile at its core. In each case the central core emission line is brighter than its surroundings. Moreover, the asymmetric line profile becomes less distinct as the distance from the central core and filament increases. The line profiles of the embedded collapsing cores can therefore be distinguished from their surroundings. 

Along the filament the line profiles have a greater blue asymmetry than in the gas outside the filament. This is particularly clear in Filament C which forms from colliding gas flows. As shown by \citet{Myers96}, a plane of two converging gas fronts also gives a blue asymmetric profile. In our filament the converging gas sheets first bring the filament together and then as the densities increase gravity causes the filament to collapse further. This reinforces the point made in the previous Section that the velocities within the filament as a whole also contribute strongly to the line profiles.

We also examined the line profiles as a function of time to test how transient our results were. For the case of Filament A shown in \fig \ref{surround}, a blue asymmetric profile was observed for at least a period of $\sim 2\E^4$ yr. After this period we expect outflows to start affecting our core, an effect which is not included in our models, and so we do not extend our analysis beyond this time period.

\section{Interpretation of line profiles}\label{discussion}

\subsection{Blue profiles}
First we consider the HCN line profiles classified as blue in \sect \ref{lines}. \tab \ref{maxints} shows the mean of the peak intensity of the HCN F(2-1) and \n(1-0) lines from the three embedded cores. In each of the cores there is a greater peak intensity of emission from the blue asymmetric line profiles compared to the red and ambiguous profiles. (The standard deviations of the calculated means have similar values at the extremes of their ranges, but there is still a clear trend towards higher intensities). However this is not true in the \n\ line profiles. A survey by \citet{Lee11} also found an increase in \n\ emission in cores with a blue asymmetry in contrast to that found here. We account for this difference by noting that our sample only contains dense collapsing cores, but the sample studied by \citet{Lee11} may also contain cores in a less evolved state that are not yet collapsing. The authors propose that cores that are less bright in \n\ are less dense and hence less likely to collapse. As all the cores studied here are already dense and collapsing they all have relatively high \n\ emission.

\begin{table}
	\centering
	\caption{The mean peak brightness temperature (in K) of the lines classified as blue, red and ambiguous when observed in HCN F(2-1) in \tab \ref{nblue}. We also calculate the standard deviation of the mean (not shown for the red core in Filament C as there is only one measurement.) The cores classified as blue have a greater HCN brightness temperature since they primarily contain emission from the dense core rather than the more diffuse envelope.}
		\begin{tabular}{l l c c c}
   	         \hline
	         \hline
	         Filament & Species & Blue & Red & Ambiguous\\
	         \hline
	         A  & HCN &  $5.92\pm1.84$ & $4.54\pm1.16$ & $4.07\pm1.53$ \\ 
	         A  & \n & $1.14\pm0.32$ & $1.42\pm0.41$ & $1.82\pm0.64$ \\ 
		 \hline	         
	         B  & HCN & $4.22\pm1.07$ & $2.32\pm0.51$ & $2.48\pm1.07$ \\
	         B  & \n &$1.09\pm0.22$  & $1.14\pm0.05$ & $1.22\pm0.20$ \\
		\hline
		 C & HCN & $5.51\pm0.89$ & $2.85$ & $3.44\pm1.63$  \\
	         C  & \n & $1.20\pm0.36$ & $1.94$ & $1.25\pm0.40$ \\
		\hline
		\end{tabular}
	\label{maxints}
\end{table}

The trend of brighter peak temperatures in the optically thick blue profiles can be understood by considering the origin of the emitting gas. In the blue cores there is typically no diffuse gas in front of the core. Therefore the optically thick emission traces the high density core and the maximum brightness temperature is higher than in the red and ambiguous cases where the lines only trace the lower density gas. However, in the optically thin species the emission from the core is always visible and the foreground material represents an extra contribution to the total brightness. Therefore for an optically thin species the brightness can actually be higher in the red and asymmetric cases.

We suggest that this may be a useful test in surveys of embedded cores. If the peak brightness is systematically higher across the survey in cores with a blue asymmetric line profile, but these identified cores do not show the same trend in their \n\ profile, it is possible that some of the cores without a blue asymmetry are also collapsing.

\subsection{Red profiles}
Next, let us consider the existence of red asymmetric line profiles in our sample of collapsing cores. The red HCN line profile seen from Filament A at [inc=180\degree, $\phi=0$\degree] in \fig \ref{sightlines} is an interesting case as it is the result of a one sided accretion flow onto the core. As the accretion is only from one direction, the blue peak is higher when viewed from [inc=0\degree, $\phi=0$\degree] and the red peak is higher when viewed from [inc=180\degree, $\phi=0$\degree].

A one sided accretion flow results when an elongated core, such as those seen in \citet{Smith11a}, is more bound on one side than the other. Such a situation arises when a dense region of a filament becomes gravitationally unstable and, due to there being no pressure support until the formation of a protostar, the original geometry of the dense gas is maintained. Gravitational focussing encourages collapse at the edges of non-spherical objects \citep{Hartmann07} and so the densest point of the core is frequently offset from the centre. The densest side of the core collapses quickly to the centre of potential, but the other side of the core is accreted more slowly leading to one sided accretion onto the dense region.

Another effect that can cause a red profile is random outward motions in the gas surrounding the core. As previously discussed, the peak brightness temperature of such profiles should be lower as the emission is coming from more diffuse gas. However, when the red asymmetry originates from a one sided accretion flow the brightness of the line is almost equal to that of the blue profiles because the emission originates from dense core gas.

Red asymmetric profiles are seen in many observations of dense cores, for instance in the \citet{Gregersen97} survey of Class 0 cores, three out out of twenty-seven cores had a red asymmetry when observed in HCO$^+$. In isolated cores surrounded by a hot confining medium, this line profile is thought to indicate expansion or oscillation \citep[e.g][]{Lada03b}. However, the analysis of Filament A has allowed us to add a new potential explanation to this list. In cores embedded within filaments a red asymmetric optically thick line profile may be due to one-sided accretion onto the core.

\subsection{Ambiguous profiles}
Lines with no clear asymmetry are the hardest to interpret. In these cases turbulent line-widths within the filament can obscure optically thick core emission at the same velocity, and the profile appears identical to what would be expected if the central core were in equilibrium. As we showed in \sect \ref{velocities}, turbulent motions are ubiquitous within filamentary structures as a result of their formation from shocks and shear flows. Consequently, the filaments in which the cores are embedded have a non-negligible velocity dispersion which can hide signatures of collapse from the optically thick lines of the central embedded core. Accordingly, a greater proportion of cores embedded within filaments may be collapsing than observations initially suggest.

In the line profiles examined here, in more than $ 50\%$ of cases the collapsing cores do not produce a blue asymmetric profile due to obscuration by their parent filament. If this were true for cores embedded within filaments in general, then a factor of 2.5 more cores may be collapsing than would be indicated by their line profiles alone. Such a substantial increase would suggest that dense cores are more unstable than previously thought.

As discussed in \sect \ref{lines}, the CS line profiles are both harder to interpret and are less reliable than the HCN observations. The vast majority of the CS emission in our models originates from the filament envelope, partly as a consequence of freeze-out effects. Due to the turbulent motions of the filament, the line profiles just trace the part of the filament along the line of sight. Any observational analysis of optically thick line profiles thought to originate from filamentary envelopes must take into account that information is only being obtained for one sight-line and that another sight-line is likely to give different results.

\subsection{Comparison to Observations}

\tab \ref{obs_comparison} outlines the numbers of blue and red line profiles found in surveys of low mass cores.  None of the surveys find the majority of the observed cores to have a blue asymmetry, but instead most find a fraction of above 30\%. Only \citet{Gregersen00} clearly excluded red asymmetries from their sample, and \citet{Sohn07} found 22\% of their cores had a red asymmetry in at least two line transitions.

\begin{table*}
	\centering
	\caption{A comparison of surveys of infall in low mass cores from the literature. As a note of caution not all surveys classify cores as `blue' using the same method. In these surveys we classify cores as blue where multiple components had a blue $\delta V$ value. When no comparable $\delta V$ values were given we used the number of infall candidates. (* Red profiles statistics were not discussed in this paper.)}
		\begin{tabular}{l c c c c}
   	         \hline
	         \hline
	         Survey & Species & No. Cores & Blue & Red \\
	         \hline
  		 \citet{Gregersen97} &  HCO$^+$ & 23 & 39\% & 13\%\\ %class 0
		 \citet{Gregersen00} &  HCO$^+$ & 17 & 35\% & 0\%\\ %starless
		 \citet{Lee99} & CS & 69 & 29\% &  4\%\\%starless
		 %\citet{Lee01} & CS & 53 & 19-36\% & 7\% \\%starless
		 \citet{Mardones97} & H$_2$CO,CS & 47 & 32\% &*\\ %class0,1
		 \citet{Andre07} & CS, HCO$^+$& 25 & 24-64\% & *\\
		 \citet{Sohn07} & HCN & 64 & 43\% &22\% \\
		 \hline
		 This Work\\	       
		 \hline  
	         By shape & HCN & 42 & 36\% & 17\% \\
	         By $\delta V$ & HCN & 42 & 48\% & 31\%\\
	         By $\delta V$ & CS & 42 & 38\% & 33\%\\
		\hline
		\end{tabular}
	\label{obs_comparison}
\end{table*}

When comparing to the observations several factors must be considered. The species used vary between the surveys and we have no HCO$^+$ measurements in this work. Each core in the surveys is by necessity observed from only a single viewing angle, and it is not stated in the observations whether any of the cores are embedded within filaments. However, \textit{Herschel} has revealed that many star-forming regions contain filaments which were previously not cleanly detected due to `chopping' filtering out background structure within the observed cloud. It is therefore likely that some fraction of the cores studied by these authors reside in more complex filamentary environments than was previously realised.

We also note that caution is in order when interpreting the comparison to observations as various selection effects can be hard to account for in the observations compared to our simulations. Many of the observational studies considered here are biased towards cores previously suspected to have infall motions. However our theoretical study is also biased as we \textit{only} contain cores that are unambiguously known to be collapsing as they go on to form stars. (The aim of our paper being to find \textit{if} a core is collapsing, how often this can be deduced from the line profile when the core is embedded in a filament within a star forming cluster.) Given this fact, biased observational studies may actually the fairest comparison.

The observations of \citet{Lee99} primarily used CS which we find to not be a reliable tracer of collapse within our embedded cores. These authors found 29\% of their cores to have blue line profiles, whereas our results were very ambiguous. There are several possible reasons for this discrepancy. Firstly not all of the \citet{Lee99} cores are in filaments, some are more isolated cores, where the envelope of the core is indeed collapsing and more easily detected. \citet{Lee11} revisited some of their sampled cores to examine their environments and found that the majority (17/35) were associated with filaments as considered here, a few cores were truly isolated (4/35) and the remainder (14/35) were surrounded by the cloud or at a cloud edge. Secondly our filaments are dense objects where CS is frozen out in the core, in less dense filament CS may be a better tracer. In observational studies it is frequently found that line profiles from different tracers vary widely \citep[e.g.][]{Shinnaga04}, in a similar manner to the theoretical work presented here.

The findings of \citet{Sohn07} are the most directly comparable to this work as they focus on HCN observations of dense cores. Their selected cores were nearby bright sources which were dense ($n\sim10^4-10^5$ \cmc), compact ($r\sim0.05-0.35$ pc) and had narrow \n\ linewidths ($\delta v \sim 0.2-0.4$ \kms). The surveyed cores, therefore, have similar physical properties to those studied here. \citet{Sohn07} has the highest proportion of both red and blue line asymmetries of the surveys shown in \tab \ref{obs_comparison}. In our models optical depth effects in the filament obscured the collapse signature from the central core in over 50\% of cases. If the same effects were present in the \citet{Sohn07} survey, it is within the bounds of possibility that the majority of their sampled cores were collapsing.

Regardless of the species the root cause of the variability with viewing angle is the intrinsic variation of the density and velocity fields with viewing angle. In our study the effect is most pronounced in CS emission as it was frozen out in the core, but there is also a high degree of variability within the HCN line profiles. Therefore we expect viewing angle effects to influence the observed line profiles of all tracer species, not just the ones studied here. This leads to the possibility that all the surveys shown in \tab \ref{obs_comparison} underestimated the number of collapsing cores, and dense cores may be more unstable objects than previously thought.

\section{Conclusions}\label{conclusions}
In this paper we consider three filaments with embedded collapsing cores from a large scale GMC simulation and perform radiative transfer calculations to model the line profiles. The filaments and their embedded cores form dynamically from bulk flows within the molecular cloud. The emission from each core is modelled using three species: optically thin \n(1-0), and optically thick CS(2-1) and HCN(1-0). Each model is observed from multiple viewing angles using a Gaussian beam through the core centre. Our main conclusions are as follows.
\begin{enumerate}
\item The velocity fields within the filaments are disordered and contain turbulent velocities. The velocities are imprinted on the filament as a result of its formation from turbulent converging flows and shocks.
\item The optically thin \n(1-0) linewidths are roughly sonic, as shown in \tab \ref{linewidths}.
\item The optically thick line profiles from the embedded cores are highly variable with viewing angle, as shown in Figures \ref{sightlines} and \ref{sightlines_cs}.
 \item HCN lines, particularly HCN F(2-1) observations are more reliable indicators of core collapse than CS(2-1). In our simulated dense dynamical filaments CS(2-1) only traces the filamentary envelope around the core within which there is no systematic collapse in our models. 
\item More than 50\% of cases show no blue-asymmetric collapse profile, due to emission from the turbulent filament surrounding the core. This effect could lead to an underestimation of the true number of collapsing cores in observational surveys.
\item Sight-lines which trace the collapsing core are distinguishable from sight-lines which mainly trace the filament, as they have a systematically higher HCN peak intensity than the rest of the sample. \n\ emission does not show such a trend.
\item In many sight-lines, a red asymmetric line profile arises from the embedded collapsing cores. This is either due to interference from the filamentary envelope, or one sided accretion onto the core.
\end{enumerate}
The data used for this work is publicly available by request from the first author at the given correspondence e-mail address.

\section*{Acknowledgements}
We are particularly indebted to Cornelius Dullemond author of RADMC-3D, without which this paper would not have been possible. We also wish to thank Paul Clark, James DiFrancesco, Neil Evans, Simon Glover, Alyssa Goodman, Ralf Launhardt and Eve Ostriker for useful discussions. We acknowledge financial support from a number of sources: a Frontier grant of Heidelberg University sponsored by the German Excellence Initiative; and NSF grants AST-0708795 and AST-1009928. In addition, we are grateful for subsidies from the German Bundesministerium f\"{u}r Bildung und Forschung via the ASTRONET project STAR FORMAT (grant 05A09VHA) as well as from the Deutsche Forschungsgemeinschaft (DFG) under grants no.\ KL 1358/10, and KL 1358/11 and via the SFB 881 `The Milky Way System«.

\bibliography{./Bibliography}

\begin{thebibliography}{75}
\expandafter\ifx\csname natexlab\endcsname\relax\def\natexlab#1{#1}\fi

\bibitem[{{Aikawa} {et~al.}(2005){Aikawa}, {Herbst}, {Roberts}, \&
  {Caselli}}]{Aikawa05}
{Aikawa}, Y., {Herbst}, E., {Roberts}, H., \& {Caselli}, P. 2005, \apj, 620,
  330

\bibitem[{{Andr{\'e}} {et~al.}(2007){Andr{\'e}}, {Belloche}, {Motte}, \&
  {Peretto}}]{Andre07}
{Andr{\'e}}, P., {Belloche}, A., {Motte}, F., \& {Peretto}, N. 2007, \aap, 472,
  519

\bibitem[{{Andr{\'e}} {et~al.}(2010){Andr{\'e}}, {Men'shchikov}, {Bontemps},
  {K{\"o}nyves}, {Motte}, {Schneider}, {Didelon}, {Minier}, {Saraceno},
  {Ward-Thompson}, {di Francesco}, {et~al.}}]{Andre10}
{Andr{\'e}}, P., {Men'shchikov}, A., {Bontemps}, S., {K{\"o}nyves}, V.,
  {Motte}, F., {Schneider}, N., {Didelon}, P., {Minier}, V., {Saraceno}, P.,
  {Ward-Thompson}, D., {di Francesco}, J., {et~al.} 2010, \aap, 518, L102+

\bibitem[{{Arzoumanian} {et~al.}(2011){Arzoumanian}, {Andr{\'e}}, {Didelon},
  {K{\"o}nyves}, {Schneider}, {Men'shchikov}, {Sousbie}, \&
  {Zavagno}}]{Arzoumanian11}
{Arzoumanian}, D., {Andr{\'e}}, P., {Didelon}, P., {K{\"o}nyves}, V.,
  {Schneider}, N., {Men'shchikov}, A., {Sousbie}, T., \& {Zavagno}, A., e.~a.
  2011, \aap, 529, L6+

\bibitem[{{Ballesteros-Paredes} {et~al.}(2003){Ballesteros-Paredes}, {Klessen},
  \& {V{\'a}zquez-Semadeni}}]{Ballesteros-Paredes03}
{Ballesteros-Paredes}, J., {Klessen}, R.~S., \& {V{\'a}zquez-Semadeni}, E.
  2003, \apj, 592, 188

\bibitem[{{Banerjee} {et~al.}(2009){Banerjee}, {V{\'a}zquez-Semadeni},
  {Hennebelle}, \& {Klessen}}]{Banerjee09}
{Banerjee}, R., {V{\'a}zquez-Semadeni}, E., {Hennebelle}, P., \& {Klessen},
  R.~S. 2009, \mnras, 398, 1082

\bibitem[{{Bergin} \& {Tafalla}(2007)}]{Bergin07}
{Bergin}, E.~A. \& {Tafalla}, M. 2007, \araa, 45, 339

\bibitem[{{Boldyrev} {et~al.}(2002){Boldyrev}, {Nordlund}, \&
  {Padoan}}]{Boldyrev02}
{Boldyrev}, S., {Nordlund}, {\AA}., \& {Padoan}, P. 2002, Physical Review
  Letters, 89, 031102

\bibitem[{{Bontemps} {et~al.}(2010){Bontemps}, {Motte}, {Csengeri}, \&
  {Schneider}}]{Bontemps10}
{Bontemps}, S., {Motte}, F., {Csengeri}, T., \& {Schneider}, N. 2010, \aap,
  524, A18+

\bibitem[{{Choi} {et~al.}(1995){Choi}, {Evans}, {Gregersen}, \&
  {Wang}}]{Choi95}
{Choi}, M., {Evans}, II, N.~J., {Gregersen}, E.~M., \& {Wang}, Y. 1995, \apj,
  448, 742

\bibitem[{{Daniel} {et~al.}(2005){Daniel}, {Dubernet}, {Meuwly}, {Cernicharo},
  \& {Pagani}}]{Daniel05}
{Daniel}, F., {Dubernet}, M.-L., {Meuwly}, M., {Cernicharo}, J., \& {Pagani},
  L. 2005, \mnras, 363, 1083

\bibitem[{{Dumouchel} {et~al.}(2010){Dumouchel}, {Faure}, \&
  {Lique}}]{Dumouchel10}
{Dumouchel}, F., {Faure}, A., \& {Lique}, F. 2010, \mnras, 406, 2488

\bibitem[{{Evans}(1999)}]{Evans99}
{Evans}, II, N.~J. 1999, \araa, 37, 311

\bibitem[{{Fuller} {et~al.}(2005){Fuller}, {Williams}, \&
  {Sridharan}}]{Fuller05}
{Fuller}, G.~A., {Williams}, S.~J., \& {Sridharan}, T.~K. 2005, \aap, 442, 949

\bibitem[{{Goodman} {et~al.}(1998){Goodman}, {Barranco}, {Wilner}, \&
  {Heyer}}]{Goodman98}
{Goodman}, A.~A., {Barranco}, J.~A., {Wilner}, D.~J., \& {Heyer}, M.~H. 1998,
  \apj, 504, 223

\bibitem[{{Green} \& {Thaddeus}(1974)}]{Green74}
{Green}, S. \& {Thaddeus}, P. 1974, \apj, 191, 653

\bibitem[{{Gregersen} \& {Evans}(2000)}]{Gregersen00}
{Gregersen}, E.~M. \& {Evans}, II, N.~J. 2000, \apj, 538, 260

\bibitem[{{Gregersen} {et~al.}(1997){Gregersen}, {Evans}, {Zhou}, \&
  {Choi}}]{Gregersen97}
{Gregersen}, E.~M., {Evans}, II, N.~J., {Zhou}, S., \& {Choi}, M. 1997, \apj,
  484, 256

\bibitem[{{Hacar} \& {Tafalla}(2011)}]{Hacar11}
{Hacar}, A. \& {Tafalla}, M. 2011, \aap, 533, A34+

\bibitem[{{Hartmann} \& {Burkert}(2007)}]{Hartmann07}
{Hartmann}, L. \& {Burkert}, A. 2007, \apj, 654, 988

\bibitem[{{Hennebelle}(2003)}]{Hennebelle03}
{Hennebelle}, P. 2003, \aap, 397, 381

\bibitem[{{Hennebelle} \& {Chabrier}(2008)}]{Hennebelle08}
{Hennebelle}, P. \& {Chabrier}, G. 2008, ArXiv e-prints, 805

\bibitem[{{J{\o}rgensen} {et~al.}(2004){J{\o}rgensen}, {Sch{\"o}ier}, \& {van
  Dishoeck}}]{Jorgensen04}
{J{\o}rgensen}, J.~K., {Sch{\"o}ier}, F.~L., \& {van Dishoeck}, E.~F. 2004,
  \aap, 416, 603

\bibitem[{{Keto} {et~al.}(2006){Keto}, {Broderick}, {Lada}, \&
  {Narayan}}]{Keto06}
{Keto}, E., {Broderick}, A.~E., {Lada}, C.~J., \& {Narayan}, R. 2006, \apj,
  652, 1366

\bibitem[{{Kirk} {et~al.}(2007){Kirk}, {Johnstone}, \& {Tafalla}}]{Kirk07}
{Kirk}, H., {Johnstone}, D., \& {Tafalla}, M. 2007, \apj, 668, 1042

\bibitem[{{Klessen} {et~al.}(2005){Klessen}, {Ballesteros-Paredes},
  {V{\'a}zquez-Semadeni}, \& {Dur{\'a}n-Rojas}}]{Klessen05}
{Klessen}, R.~S., {Ballesteros-Paredes}, J., {V{\'a}zquez-Semadeni}, E., \&
  {Dur{\'a}n-Rojas}, C. 2005, \apj, 620, 786

\bibitem[{{Klessen} {et~al.}(2000){Klessen}, {Heitsch}, \& {Mac
  Low}}]{Klessen00b}
{Klessen}, R.~S., {Heitsch}, F., \& {Mac Low}, M. 2000, \apj, 535, 887

\bibitem[{{Klessen} \& {Hennebelle}(2010)}]{Klessen10}
{Klessen}, R.~S. \& {Hennebelle}, P. 2010, \aap, 520, A17+

\bibitem[{{Lada} {et~al.}(2003){Lada}, {Bergin}, {Alves}, \& {Huard}}]{Lada03b}
{Lada}, C.~J., {Bergin}, E.~A., {Alves}, J.~F., \& {Huard}, T.~L. 2003, \apj,
  586, 286

\bibitem[{{Larson}(1969)}]{Larson69}
{Larson}, R.~B. 1969, \mnras, 145, 271

\bibitem[{{Lee} \& {Myers}(2011)}]{Lee11}
{Lee}, C.~W. \& {Myers}, P.~C. 2011, \apj, 734, 60

\bibitem[{{Lee} {et~al.}(1999){Lee}, {Myers}, \& {Tafalla}}]{Lee99}
{Lee}, C.~W., {Myers}, P.~C., \& {Tafalla}, M. 1999, \apj, 526, 788

\bibitem[{{Lee} {et~al.}(2004){Lee}, {Bergin}, \& {Evans}}]{LeeJ04}
{Lee}, J.-E., {Bergin}, E.~A., \& {Evans}, II, N.~J. 2004, \apj, 617, 360

\bibitem[{{Mardones} {et~al.}(1997){Mardones}, {Myers}, {Tafalla}, {Wilner},
  {Bachiller}, \& {Garay}}]{Mardones97}
{Mardones}, D., {Myers}, P.~C., {Tafalla}, M., {Wilner}, D.~J., {Bachiller},
  R., \& {Garay}, G. 1997, \apj, 489, 719

\bibitem[{{Men'shchikov} {et~al.}(2010){Men'shchikov}, {Andr{\'e}}, {Didelon},
  {K{\"o}nyves}, {Schneider}, {Motte}, {Bontemps}, {et~al.}}]{Menshchikov10}
{Men'shchikov}, A., {Andr{\'e}}, P., {Didelon}, P., {K{\"o}nyves}, V.,
  {Schneider}, N., {Motte}, F., {Bontemps}, S., {et~al.} 2010, \aap, 518, L103+

\bibitem[{{Myers} {et~al.}(1996){Myers}, {Mardones}, {Tafalla}, {Williams}, \&
  {Wilner}}]{Myers96}
{Myers}, P.~C., {Mardones}, D., {Tafalla}, M., {Williams}, J.~P., \& {Wilner},
  D.~J. 1996, \apjl, 465, L133+

\bibitem[{{Ossenkopf}(1997)}]{Ossenkopf97}
{Ossenkopf}, V. 1997, \na, 2, 365

\bibitem[{{Ostriker} {et~al.}(2001){Ostriker}, {Stone}, \&
  {Gammie}}]{Ostriker01}
{Ostriker}, E.~C., {Stone}, J.~M., \& {Gammie}, C.~F. 2001, \apj, 546, 980

\bibitem[{{Ostriker}(1964)}]{Ostriker64}
{Ostriker}, J. 1964, \apj, 140, 1529

\bibitem[{{Park} \& {Hong}(1998)}]{Park98}
{Park}, Y.-S. \& {Hong}, S.~S. 1998, \apj, 494, 605

\bibitem[{{Penston}(1969)}]{Penston69}
{Penston}, M.~V. 1969, \mnras, 144, 425

\bibitem[{{Pichardo} {et~al.}(2000){Pichardo}, {V{\'a}zquez-Semadeni}, {Gazol},
  {Passot}, \& {Ballesteros-Paredes}}]{Pichardo00}
{Pichardo}, B., {V{\'a}zquez-Semadeni}, E., {Gazol}, A., {Passot}, T., \&
  {Ballesteros-Paredes}, J. 2000, \apj, 532, 353

\bibitem[{{Pineda} {et~al.}(2011){Pineda}, {Goodman}, {Arce}, {Caselli},
  {Longmore}, \& {Corder}}]{Pineda11}
{Pineda}, J.~E., {Goodman}, A.~A., {Arce}, H.~G., {Caselli}, P., {Longmore},
  S., \& {Corder}, S. 2011, \apjl, 739, L2+

\bibitem[{{Rawlings} \& {Yates}(2001)}]{Rawlings01}
{Rawlings}, J.~M.~C. \& {Yates}, J.~A. 2001, \mnras, 326, 1423

\bibitem[{{Robitaille} {et~al.}(2006){Robitaille}, {Whitney}, {Indebetouw},
  {Wood}, \& {Denzmore}}]{Robitaille06}
{Robitaille}, T.~P., {Whitney}, B.~A., {Indebetouw}, R., {Wood}, K., \&
  {Denzmore}, P. 2006, \apjs, 167, 256

\bibitem[{{Schmidt} {et~al.}(2008){Schmidt}, {Federrath}, \&
  {Klessen}}]{Schmidt08}
{Schmidt}, W., {Federrath}, C., \& {Klessen}, R. 2008, Physical Review Letters,
  101, 194505

\bibitem[{{Sch{\"o}ier} {et~al.}(2005){Sch{\"o}ier}, {van der Tak}, {van
  Dishoeck}, \& {Black}}]{Schoier05}
{Sch{\"o}ier}, F.~L., {van der Tak}, F.~F.~S., {van Dishoeck}, E.~F., \&
  {Black}, J.~H. 2005, \aap, 432, 369

\bibitem[{{She} \& {Leveque}(1994)}]{She94}
{She}, Z.-S. \& {Leveque}, E. 1994, Physical Review Letters, 72, 336

\bibitem[{{Shetty} {et~al.}(2010){Shetty}, {Collins}, {Kauffmann}, {Goodman},
  {Rosolowsky}, \& {Norman}}]{Shetty10}
{Shetty}, R., {Collins}, D.~C., {Kauffmann}, J., {Goodman}, A.~A.,
  {Rosolowsky}, E.~W., \& {Norman}, M.~L. 2010, \apj, 712, 1049

\bibitem[{{Shetty} {et~al.}(2011{\natexlab{a}}){Shetty}, {Glover}, {Dullemond},
  \& {Klessen}}]{Shetty11}
{Shetty}, R., {Glover}, S.~C., {Dullemond}, C.~P., \& {Klessen}, R.~S.
  2011{\natexlab{a}}, \mnras, 412, 1686

\bibitem[{{Shetty} {et~al.}(2011{\natexlab{b}}){Shetty}, {Glover}, {Dullemond},
  {Ostriker}, {Harris}, \& {Klessen}}]{Shetty11b}
{Shetty}, R., {Glover}, S.~C., {Dullemond}, C.~P., {Ostriker}, E.~C., {Harris},
  A.~I., \& {Klessen}, R.~S. 2011{\natexlab{b}}, \mnras, 415, 3253

\bibitem[{{Shinnaga} {et~al.}(2004){Shinnaga}, {Ohashi}, {Lee}, \&
  {Moriarty-Schieven}}]{Shinnaga04}
{Shinnaga}, H., {Ohashi}, N., {Lee}, S.-W., \& {Moriarty-Schieven}, G.~H. 2004,
  \apj, 601, 962

\bibitem[{{Shu}(1977)}]{Shu77}
{Shu}, F.~H. 1977, \apj, 214, 488

\bibitem[{{Smith} {et~al.}(2009{\natexlab{a}}){Smith}, {Clark}, \&
  {Bonnell}}]{Smith09a}
{Smith}, R.~J., {Clark}, P.~C., \& {Bonnell}, I.~A. 2009{\natexlab{a}}, \mnras,
  396, 830

\bibitem[{{Smith} {et~al.}(2011){Smith}, {Glover}, {Bonnell}, {Clark}, \&
  {Klessen}}]{Smith11a}
{Smith}, R.~J., {Glover}, S.~C.~O., {Bonnell}, I.~A., {Clark}, P.~C., \&
  {Klessen}, R.~S. 2011, \mnras, 411, 1354

\bibitem[{{Smith} {et~al.}(2009{\natexlab{b}}){Smith}, {Longmore}, \&
  {Bonnell}}]{Smith09b}
{Smith}, R.~J., {Longmore}, S., \& {Bonnell}, I. 2009{\natexlab{b}}, \mnras,
  400, 1775

\bibitem[{{Sobolev}(1957)}]{Sobolev57}
{Sobolev}, V.~V. 1957, \sovast, 1, 678

\bibitem[{{Sohn} {et~al.}(2007){Sohn}, {Lee}, {Park}, {Lee}, {Myers}, \&
  {Lee}}]{Sohn07}
{Sohn}, J., {Lee}, C.~W., {Park}, Y.-S., {Lee}, H.~M., {Myers}, P.~C., \&
  {Lee}, Y. 2007, \apj, 664, 928

\bibitem[{{Stahler} \& {Yen}(2010)}]{Stahler10}
{Stahler}, S.~W. \& {Yen}, J.~J. 2010, \mnras, 407, 2434

\bibitem[{{Stutz} {et~al.}(2009){Stutz}, {Rieke}, {Bieging}, {Balog},
  {Heitsch}, {Kang}, {Peters}, {Shirley}, \& {Werner}}]{Stutz09}
{Stutz}, A.~M., {Rieke}, G.~H., {Bieging}, J.~H., {Balog}, Z., {Heitsch}, F.,
  {Kang}, M., {Peters}, W.~L., {Shirley}, Y.~L., \& {Werner}, M.~W. 2009, \apj,
  707, 137

\bibitem[{{Stutz} {et~al.}(2008){Stutz}, {Rubin}, {Werner}, {Rieke}, {Bieging},
  {Keene}, {Kang}, {Shirley}, {Su}, {Velusamy}, \& {Wilner}}]{Stutz08}
{Stutz}, A.~M., {Rubin}, M., {Werner}, M.~W., {Rieke}, G.~H., {Bieging}, J.~H.,
  {Keene}, J., {Kang}, M., {Shirley}, Y.~L., {Su}, K.~Y.~L., {Velusamy}, T., \&
  {Wilner}, D.~J. 2008, \apj, 687, 389

\bibitem[{{Tafalla} {et~al.}(2002){Tafalla}, {Myers}, {Caselli}, {Walmsley}, \&
  {Comito}}]{Tafalla02}
{Tafalla}, M., {Myers}, P.~C., {Caselli}, P., {Walmsley}, C.~M., \& {Comito},
  C. 2002, \apj, 569, 815

\bibitem[{{Tilley} \& {Pudritz}(2003)}]{Tilley03}
{Tilley}, D.~A. \& {Pudritz}, R.~E. 2003, \apj, 593, 426

\bibitem[{{Tobin} {et~al.}(2010){Tobin}, {Hartmann}, {Looney}, \&
  {Chiang}}]{Tobin10}
{Tobin}, J.~J., {Hartmann}, L., {Looney}, L.~W., \& {Chiang}, H. 2010, \apj,
  712, 1010

\bibitem[{{Tsamis} {et~al.}(2008){Tsamis}, {Rawlings}, {Yates}, \&
  {Viti}}]{Tsamis08}
{Tsamis}, Y.~G., {Rawlings}, J.~M.~C., {Yates}, J.~A., \& {Viti}, S. 2008,
  \mnras, 388, 898

\bibitem[{{Turner} {et~al.}(1992){Turner}, {Chan}, {Green}, \&
  {Lubowich}}]{Turner92}
{Turner}, B.~E., {Chan}, K.-W., {Green}, S., \& {Lubowich}, D.~A. 1992, \apj,
  399, 114

\bibitem[{{Walker} {et~al.}(1994){Walker}, {Narayanan}, \& {Boss}}]{Walker94}
{Walker}, C.~K., {Narayanan}, G., \& {Boss}, A.~P. 1994, \apj, 431, 767

\bibitem[{{Ward-Thompson} {et~al.}(2010){Ward-Thompson}, {Kirk}, {Andr{\'e}},
  {Saraceno}, {Didelon}, {K{\"o}nyves}, {Schneider},
  {et~al.}}]{Ward-Thompson10}
{Ward-Thompson}, D., {Kirk}, J.~M., {Andr{\'e}}, P., {Saraceno}, P., {Didelon},
  P., {K{\"o}nyves}, V., {Schneider}, N., {et~al.} 2010, \aap, 518, L92+

\bibitem[{{Wilner} {et~al.}(2000){Wilner}, {Myers}, {Mardones}, \&
  {Tafalla}}]{Wilner00}
{Wilner}, D.~J., {Myers}, P.~C., {Mardones}, D., \& {Tafalla}, M. 2000, \apjl,
  544, L69

\bibitem[{{Wu} \& {Evans}(2003)}]{Wu03}
{Wu}, J. \& {Evans}, II, N.~J. 2003, \apjl, 592, L79

\bibitem[{{Zhou}(1992)}]{Zhou92}
{Zhou}, S. 1992, \apj, 394, 204

\bibitem[{{Zhou} {et~al.}(1991){Zhou}, {Evans}, {Guesten}, {Mundy}, \&
  {Kutner}}]{Zhou91}
{Zhou}, S., {Evans}, II, N.~J., {Guesten}, R., {Mundy}, L.~G., \& {Kutner},
  M.~L. 1991, \apj, 372, 518

\bibitem[{{Zhou} {et~al.}(1993){Zhou}, {Evans}, {Koempe}, \&
  {Walmsley}}]{Zhou93}
{Zhou}, S., {Evans}, II, N.~J., {Koempe}, C., \& {Walmsley}, C.~M. 1993, \apj,
  404, 232

\end{thebibliography}

%% If you are not including electonic art with your submission, you may
%% mark up your captions using the \figcaption command. See the
%% No more than seven \figcaption commands are allowed per page,
%% so if you have more than seven captions, insert a \clearpage

\end{document}